\documentclass{article}
\usepackage{graphicx} 
\usepackage{pgfplots}
\usetikzlibrary{intersections}
\pgfplotsset{compat=1.17}
\usepackage{subcaption}
\usepackage{float}
\usepackage{mathtools}   
\usepackage[normalem]{ulem}
\usepackage{hyperref}
\usepackage[flushleft]{threeparttable}
\usepackage{booktabs, makecell}
\usepackage{xcolor}
\usepackage{bbm}
\usepackage{colortbl} 
\usepackage{placeins} 
\usepackage [english]{babel}
\usepackage [autostyle, english = american]{csquotes}
\MakeOuterQuote{"}
\usepackage{authblk}
\usepackage{amssymb}
\newcommand{\beginsupplement}{
  \setcounter{table}{0}
  \renewcommand{\thetable}{A\arabic{table}}
  \setcounter{figure}{0}
  \renewcommand{\thefigure}{A\arabic{figure}}
  \setcounter{equation}{0}
  \renewcommand{\theequation}{A\arabic{equation}}
}
\graphicspath{{figs/}}

\title{Pregnancy as a dynamical paradox: \\robustness, control and birth onset}
\author[1]{Giuseppe M. Ferro}
\author[2]{Andrea Somazzi}
\author[3]{Didier Sornette\thanks{Corresponding author: \href{mailto:dsornette@ethz.ch}{dsornette@ethz.ch}}}

\affil[1]{Center for Statistics \& Machine Learning, Princeton University, Princeton, USA}
\affil[2]{IMT School for Advanced Studies, Piazza S. Francesco 19, 55100 Lucca, Italy}
\affil[3]{Institute of Risk Analysis, Prediction and Management (Risks-X)
Southern University of Science and Technology (SUSTech)
Shenzhen, 518055, China}

\begin{document}

\maketitle
\maketitle

{
\vspace{-1.2cm}
\begin{center}\footnotesize \textbf{Authors’ Contributions}\end{center}
\vspace{-0.3cm}
\footnotesize
G.M.F. and A.S. contributed equally to this work.
G.M.F., A.S. and D.S. jointly conceived the study and designed the research.
G.M.F. and A.S carried out the simulations and analyzed the results. 
All authors contributed to writing the manuscript and approved the final version.
}

\begin{abstract}

The timing of human labor is among the most critical determinants of neonatal survival, yet the mechanisms that govern the transition from uterine quiescence to coordinated contractions remain elusive. Here we present a dynamical-systems framework that models the pregnant uterus as a spatially extended network of electrically excitable cells regulated by sparse adaptive feedback mimicking hormonal and mechanical influences. This approach reveals how stability during gestation and sensitivity near parturition can be simultaneously maintained through the interplay of control, network structure, and noise. Our analysis shows that spontaneous contractions such as Braxton-Hicks and Alvarez waves are not epiphenomena, but functional components that reduce control effort and preserve responsiveness. Moreover, we identify preterm labor as a boundary-crossing phenomenon arising when control fails to correctly interpret early-warning signals. These results establish a unifying mechanistic theory for labor onset, yield testable predictions, and suggest new therapeutic strategies to mitigate preterm birth risk.
\end{abstract}

{\bf Significance Statement}.
Birth represents one of the most finely tuned transitions in human physiology, where even small deviations in timing can have life-threatening consequences. Preterm birth alone accounts for nearly one million deaths each year worldwide, yet the mechanisms that determine when the uterus shifts from quiescence to active labor remain unclear. By framing the pregnant uterus as a controlled network of interacting excitable cells, we uncover how stability is maintained during pregnancy while sensitivity is preserved near delivery. This perspective explains the functional role of seemingly “false” contractions and identifies failure of control as a pathway to preterm labor. The framework provides testable predictions and suggests new therapeutic strategies for reducing adverse birth outcomes.

\section{Introduction}
\label{sec:intro}
Pregnancy is a highly regulated and multifaceted biological process that enables the development of new life, characterized by a dynamic interplay between physiological robustness and adaptive flexibility within and across the maternal and fetal systems, including immunological, endocrine, and vascular domains. However, when this delicate balance is disrupted, serious complications can arise. Among gestational disorders, preterm birth remains a major clinical and societal challenge, occurring in over 10\% of pregnancies and accounting for more than a third of infant deaths \cite{martin2007births,macdorman2007trends}. Beyond early mortality, preterm birth leads to long-term health complications, including neurodevelopmental disabilities, compromised immune function, and chronic conditions that place a considerable economic burden on families and healthcare systems \cite{behrman2007preterm,KhanPetrou2015,Liangetalancet24,Huglobalburden25}. Understanding the mechanisms that maintain uterine quiescence, as well as the conditions under which they may fail, is therefore of both clinical and scientific importance.

The maintenance of uterine quiescence throughout pregnancy presents a remarkable control challenge: the system must be robust enough to prevent premature contractions, flexible enough to initiate labor at term, and efficient enough to function over months with minimal energetic expenditure. A central question we address is how the uterus sustains this delicate state of controlled inactivity, despite its intrinsic capacity for the forceful and coordinated contractions required during labor \cite{weiss2000endocrinology, vidal2022spontaneous}. This quiescence is known to be actively maintained by a sophisticated control system that suppresses synchronization among uterine myocytes \cite{weiss2000endocrinology}. Yet this control is not flawless, as evidenced by the occurrence of spontaneous non-labor contractions such as those described by Alvarez and Braxton Hicks \cite{russo2021alvarez, esgalhado2020uterine}. We interpret these events as manifestations of transient imperfections in a cost-effective regulatory system that suppresses premature activity while simultaneously minimizing energy expenditure. Furthermore, we propose that these contractions provide feedback on control adequacy, prompting adaptive modulation of suppressive strength in response to evolving physiological conditions.

Existing mathematical models of parturition have offered valuable insights but exhibit key limitations. One prominent class conceptualizes labor as a phase transition or bifurcation, implying a gradual increase in susceptibility to perturbations as term approaches \cite{Sornpapier94,benson2006endogenous, singh2012self}. However, this framework appears inconsistent with the clinical observation of remarkable robustness throughout gestation. Another modeling approach focuses on the electro-chemical-mechanical properties of the uterus and cervix \cite{fang2021anisotropic, fidalgo2022effect}, but largely omits the regulatory control mechanisms that actively sustain quiescence. More recently, machine learning models have achieved impressive predictive accuracy for labor onset based on clinical data \cite{kloska2025predicting}, yet they offer little insight into the underlying mechanisms.
Taken together, the literature leaves several fundamental puzzles unresolved. First, despite decades of clinical and biological research, preterm birth remains poorly understood at a mechanistic level: it is associated with a wide range of heterogeneous risk factors and pathological correlates, yet no unifying dynamical explanation has emerged for how and why the uterus transitions prematurely from quiescence to sustained contractions \cite{vidal2022spontaneous}. Second, pregnancy is marked by the frequent occurrence of spontaneous, non-labor contractions, such as Alvarez waves and Braxton–Hicks contractions, whose functional role remains debated. These events are often interpreted as benign or preparatory “training” of the uterine muscle, but this hypothesis lacks a clear mechanistic justification and does not explain their timing, variability, or apparent coexistence with robust suppression of labor. Third, and more broadly, pregnancy presents a robustness–flexibility paradox: the uterus must remain extraordinarily stable against premature activation for months, yet be capable of switching rapidly and decisively to coordinated, global contractions at term. Standard bifurcation-based or maturation-driven models struggle to reconcile these opposing requirements within a single framework. Clarifying how these properties can coexist and how their balance can fail is a central challenge for a dynamical theory of parturition.

In this paper, we introduce a novel modeling framework that emphasizes the critical role of the control system in pregnancy. Our model explicitly incorporates the mechanisms that maintain uterine quiescence and prevent premature labor. It predicts both Alvarez contractions, which are frequent, localized, and low-amplitude, and the less frequent but stronger Braxton-Hicks contractions, and it identifies the conditions under which these transient oscillatory behaviors can coexist. Importantly, the model conceptualizes preterm birth as a catastrophic failure of control, leading to premature, sustained uterine contractions. This framework reveals multiple pathways to preterm labor, expressed as distinct parameter trajectories within the model’s phase space.
Moreover, our model proposes that labor onset is not a bifurcation but a rapid and energetically efficient switch, characterized by an abrupt interruption of control that allows for an immediate transition from quiescence to coordinated contraction. This view aligns more closely with empirical observations and suggests novel directions for clinical intervention.

The paper is structured as follows: Section \ref{sec:methods} introduces the mathematical model of the uterus and its control. Section \ref{sec:res} presents our main findings (with additional analyses in the SI). Section \ref{sec:discussion} explores the broader implications of the model, and Section \ref{sec:conc} concludes with potential avenues for future research.

\section{Methods}
\label{sec:methods}
To build our modeling framework, we begin by revisiting and extending the approach introduced in \cite{singh2012self}. In their model, the uterine tissue is represented as a two-dimensional medium of size $L \times L$, consisting of excitable cells (uterine myocytes) interspersed with electrically passive cells such as fibroblasts and interstitial Cajal-like cells.
Each excitable cell $i\in \{0,\dots, L^2\}$  has voltage $V_i$ and effective membrane conductance $g_i$. Excitable cell $i$ is linearly coupled, with coupling strength $C_r$, to $n_{p,i}$ identical passive
cells having voltage $V_{p,i}$. In addition to the coupling between active and passive cells, each active cell is coupled linearly to its nearest neighbors with strength $D$ (passive cells are not coupled to each other). 

\begin{figure}[H]
    \centering
    
        \includegraphics[width=\textwidth]{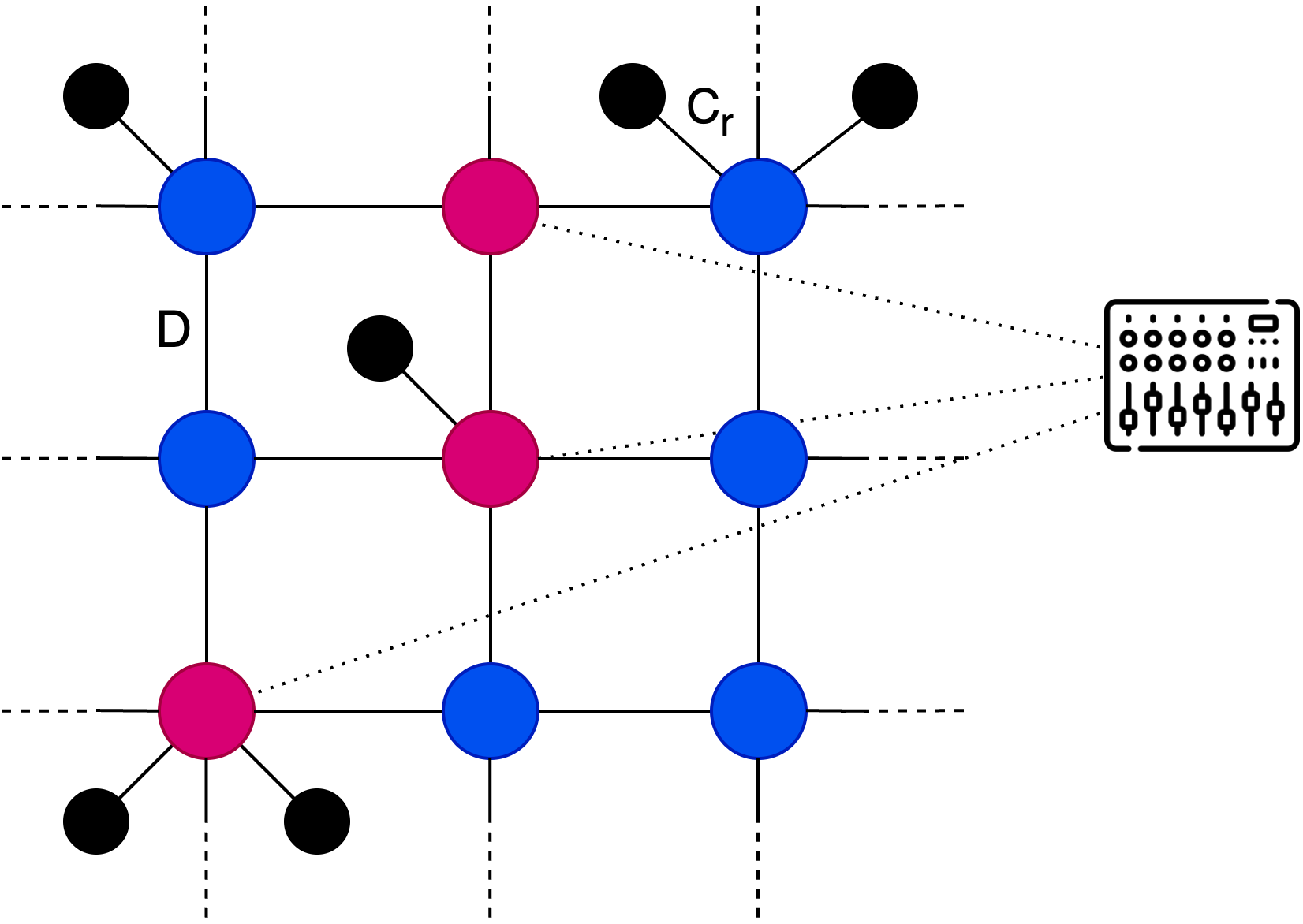}
    \caption{Schematic representation of the 2D uterine network: active cells (big circles) can be either controlled (pink) or not (blue). They are coupled to each other with strength $D$. Each active cell is connected to a random number of passive cells (small black cirles) with coupling $C_r$.}
    \label{fig:schemalattice}
\end{figure}

We generalize this model along two dimensions: 
\begin{enumerate}
    \item For a designated subset $\mathcal{C}$
of excitable cells, we introduce a linear adaptive control term $\mu_i$ for each cell $i\in \mathcal{C}$, designed to suppress its intrinsic oscillatory activity. The cardinal of $\mathcal{C}$ is $|\mathrm{\mathcal{C}}| = n_c\,L^2$ where $n_c$ is thus the fraction of excitable cells that are controlled.
    \item For each excitable cell $i$, a Gaussian noise term $\chi_{v,i}$ is introduced (see Eq. \ref{eq:sys}), in order to capture the inherent variability in biological systems 
\end{enumerate}
See Fig. \ref{fig:schemalattice} for a schematic representation of the system.
A relevant analogy comes from neuronal systems, where channel noise, referring to the stochastic opening and closing of ion channels, is often considered negligible due to the large number of channels involved. However, studies have shown that such noise can have significant physiological effects \cite{white2000channel}. Similarly, while incorporating noise into our model may initially seem like an unnecessary complication, our results demonstrate that it is essential for capturing key real-life phenomena such as Braxton-Hicks contractions. These transient, spontaneous events cannot be reproduced within a purely deterministic framework, underscoring the critical role of stochasticity in accurately modeling the physiological behavior of the uterus.

The dynamics is described by the following system of stochastic differential equations:

\begin{equation}
\label{eq:sys}
\left\{
\begin{aligned}
\frac{dV_i}{dt} \; &= \; A\,V_i\,\bigl(V_i - \omega\bigr)\,\bigl(1 - V_i\bigr)
\;-\; g_i
\;+\; n_{p,i}\,C_r\,\bigl(V_{p,i} - V_i\bigr)
\;+\; D\,(\nabla^2 V)_i
\;-\; \mu_i
\;+\; \chi_{v,i}(t)
\\[6pt]
\frac{dg_i}{dt} \; &= \; \epsilon \,\bigl(V_i - g_i\bigr)
\;
\\[6pt]
\frac{dV_{p,i}}{dt} \; &= \; K\,\bigl(V_p^{R} - V_{p,i}\bigr)-C_r\bigl(V_{p,i} - V_i\bigr)
\;
\\[6pt]
\frac{d\mu_i}{dt} \; &= \bigl[ \gamma\,\bigl(\alpha\,r_i(t-\tau)\;-\;\mu_i \bigr)
\;\bigr]\mathbbm{1}_{\mathcal{C}}(i),\quad r_i=V_i(t-\tau)-V_T~.
\end{aligned}
\right.
\end{equation}
In the first three lines of the above system of equations, parameter $A$ specifies the activation kinetics of the active cell, $ \omega$ is the excitation threshold, $\nabla^2$ is the discrete Laplace operator and $\epsilon$ dictates the recovery rate of the medium. The number $n_{p,i}$ of passive cells connected to active cell $i$  is distributed according to a Poisson distribution with mean value $f=0.7$. For the passive cell, $V_p^R$ is the resting state while $K$ characterizes how fast the system relax to $V_p^R$. The parameters $A(=3), \omega(=0.2), \epsilon(=0.08), K(=0.25), V_p^R(=1.5)$ are fixed to the values suggested by \cite{singh2012self}. 

The last line of the above system of equations describes the evolution of the control, where $r_i=V_i(t-\tau)-V_T$ is the difference between the measured voltage (with delay $\tau$) and target cell's voltage $V_T$. The indicator function $\mathbbm{1}_{\mathcal{C}}(i)$ is equal to $1$ if $i\in \mathcal{C}$ and to $0$ otherwise. We assume throughout the paper that $\mu_i(0)=0 \enspace \forall i$. The presence of the minus $\mu$ term in the r.h.s. of the equation represents the self regulation of the control, which otherwise could potentially become unrealistically  large. The trade-off between suppression of oscillation and cost of control is captured by the parameter $\alpha \geq 0$. The control evolves with a time-scale $\gamma^{-1}$. 
The noise vector is Gaussian with mean zero and correlation function $\langle \chi_{v,i}(t) \chi_{v,j}(t')\rangle=\sigma^2\delta_{ij}\delta(t-t')$.

In principle, each site can have its own control parameters $\gamma_{i}, \tau_{i}$ and $\alpha_{i}$. For simplicity, we fix the time scale $\gamma=1$ and the delay $\tau=0$ uniformly across the medium, while allowing for heterogeneity in the excitability parameters $\{\alpha_{i}\}$ to investigate their impact. We have done this by assuming that a fraction $n_c$ of excitable cells are controlled. This can be expressed alternatively by attributing randomly a parameter $\alpha_{i}$ to each excitable cell according to the following bimodal distribution:
\begin{equation}
\label{eq:bimodalcontrol}
    \rho(\alpha_{i})=n_c\delta(\alpha_{i}-\alpha)+(1-n_c)\delta(\alpha_{i})
\end{equation}
Eq. \ref{eq:bimodalcontrol} amounts to having a fraction $n_c$ of controlled sites with parameter $\alpha$. This bimodal distribution simply reflects the assumption that the fraction $n_c$ of excitable cells that are the subset $\mathcal{C}$
are subject to control, such that the excitability parameters $\alpha_{i}$ take one value for controlled cells and the value $0$ for uncontrolled ones. It is possible to generalise this distribution (\ref{eq:bimodalcontrol}) to account for a differential sensitivity of cells to control. For example, within the myometrium, single-cell RNA sequencing reveals a mosaic of varying levels of progesterone resistance \cite{ji2024exploring}, a crucial hormone used to inhibit cells contractions. We keep the form (\ref{eq:bimodalcontrol}) for simplicity.

In absence of control, for $D=0$, each sub-system of active-passive cells behaves essentially as a FitzHugh-Nagumo system, undergoing a Hopf bifurcation as a function of $n_p$ and $C_r$, thus exhibiting stable oscillations in certain parameter space (see Fig. 1 in \cite{singh2012self}). Increasing $D$, synchronized oscillatory activity emerges, leading to: i) increased correlation between cells; ii) more and more cells oscillate with the same frequency (see Figs. 2 and 3 in \cite{singh2012self}).

We have simulated the equations (\ref{eq:sys}) with a fourth-order Runge-Kutta algorithm, with a time step of $dt=0.1$, for $2^{15}$ time units.

\section{Results}
\label{sec:res}

\subsection{Control Performance}
\label{secsub:scellcontrol}

Let us start by briefly describing the behavior of a single controlled excitable cell in the absence of noise ($\chi_{v}=0$) as described by 
\begin{equation}
\label{eq:sys1dPRIMA}
\left\{
\begin{aligned}
\frac{dV}{dt} \; &= \; A\,V\,\bigl(V - \omega\bigr)\,\bigl(1 - V\bigr)
\;-\; g
\;+\; n_{p}\,C_r\,\bigl(V_{p} - V\bigr)
\;-\; \mu
\\[6pt]
\frac{dg}{dt} \; &= \; \epsilon \,\bigl(V - g\bigr)
\\[6pt]
\frac{dV_{p}}{dt} \; &= \; K\,\bigl(V_p^{R} - V_{p}\bigr)-C_r\bigl(V_{p} - V\bigr)
\\[6pt]
\frac{d\mu}{dt} \; &=  \gamma\,\bigl(\alpha\,r(t-\tau)\;-\;\mu \bigr)
\end{aligned}
\right.
\end{equation}
The parameters $C_r$($=1$) and $n_p$($=1$) are set so that: i) the system has a unique steady state; ii) the cell would spontaneously oscillate when the control is switched-off ($\alpha=0$).

For $\tau = 0$ (see Section \ref{sec:delay_control} for details about the effect of a non-zero $\tau$), the system undergoes a supercritical Hopf bifurcation at a critical control gain $\alpha_c$ (see Fig. \ref{fig:pd_single_cell}). For $\alpha > \alpha_c$, the dynamics converge to a stable fixed point $(V^*, g^*, V_p^*, \mu^*)$ satisfying
\[
 g^* = V^*, \qquad \mu^* = \alpha\,(V^*-V_T), \qquad (K+C_r)\,V_p^* = K\,V_p^R + C_r\,V^*.
\]
Using $V_p^* - V^* = \dfrac{K}{K+C_r}\,(V_p^R - V^*)$, the $dV/dt = 0$ equation becomes
\begin{equation}
\label{eq:Vstar_equation}
A\,V^*(V^* - \omega)(1 - V^*) \, - \, V^* \, + \, n_p\,\underbrace{\frac{K C_r}{K+C_r}}_{\displaystyle \tilde C_r}\,(V_p^R - V^*) \, - \, \alpha\,(V^* - V_T) \;=\; 0,
\end{equation}
This equation has a unique stable solution $V^*$ close to the prescribed target voltage $V_T$ whenever $\alpha > \alpha_c$. In this regime, oscillations are suppressed and the control term $\mu$ settles to a constant value $\mu^*$ that exactly counterbalances the cell’s intrinsic excitability, keeping the membrane potential clamped near $V_T$. 

This statement can be made quantitative as follows. Define the nonlinear term
\[
G(V) := A\,V(V-\omega)(1-V) - V + n_p\,\tilde C_r\,(V_p^R - V),
\]
and write the steady-state condition (Eq. \ref{eq:Vstar_equation}) as
\[
G(V^*) - \alpha\,(V^* - V_T) = 0.
\]
Letting $e := V^* - V_T$ and linearising $G$ around $V_T$ gives
\[
G(V_T+e) \approx G(V_T) + G'(V_T)\,e,
\]
so that
\[
 e \;=\; V^* - V_T \;\approx\; \frac{G(V_T)}{\alpha - G'(V_T)} 
 \,=\, \frac{\displaystyle A\,V_T(V_T-\omega)(1-V_T) - (1+n_p\,\tilde C_r)\,V_T + n_p\,\tilde C_r\,V_p^R}{\displaystyle \alpha - \Bigl[A\,\bigl(-3V_T^2 + 2(1+\omega)V_T - \omega\bigr) - 1 - n_p\,\tilde C_r\Bigr]}.
\]
For $\alpha > G'(V_T)$, this shows that
\[
|V^* - V_T| \;\lesssim\; \frac{|G(V_T)|}{\alpha - G'(V_T)} = \mathcal{O}\!\left(\frac{1}{\alpha}\right),
\]
where the last equality leading to $\mathcal{O}\!\left(\frac{1}{\alpha}\right)$
is valid for large $\alpha$. This implies that
the steady-state voltage approaches the target value with an error that decays inversely with $\alpha$. In particular, if $G(V_T) = 0$, then $V^* = V_T$ exactly.

\begin{figure}[htbp]
  \centering
  \begin{minipage}[t]{0.49\textwidth}
    \centering
    \subcaptionbox{\label{figDS:main}}[
      \linewidth]{\includegraphics[width=\linewidth]{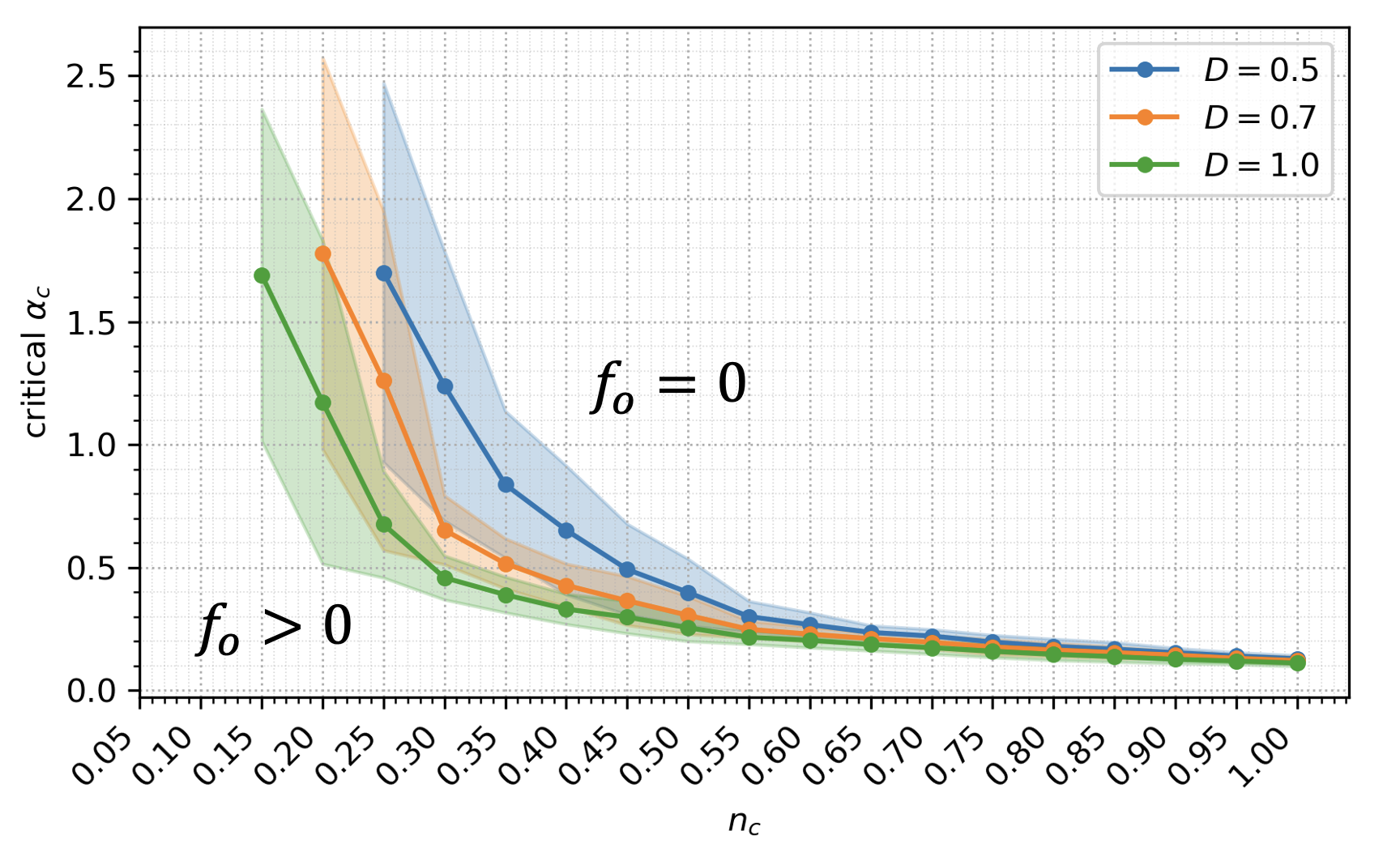}}\par\vspace{0.8em}
    
    \subcaptionbox{\label{figDS:examples}}[
      \linewidth]{
      \begin{minipage}{\linewidth}
        \centering
        \includegraphics[width=0.5\linewidth]{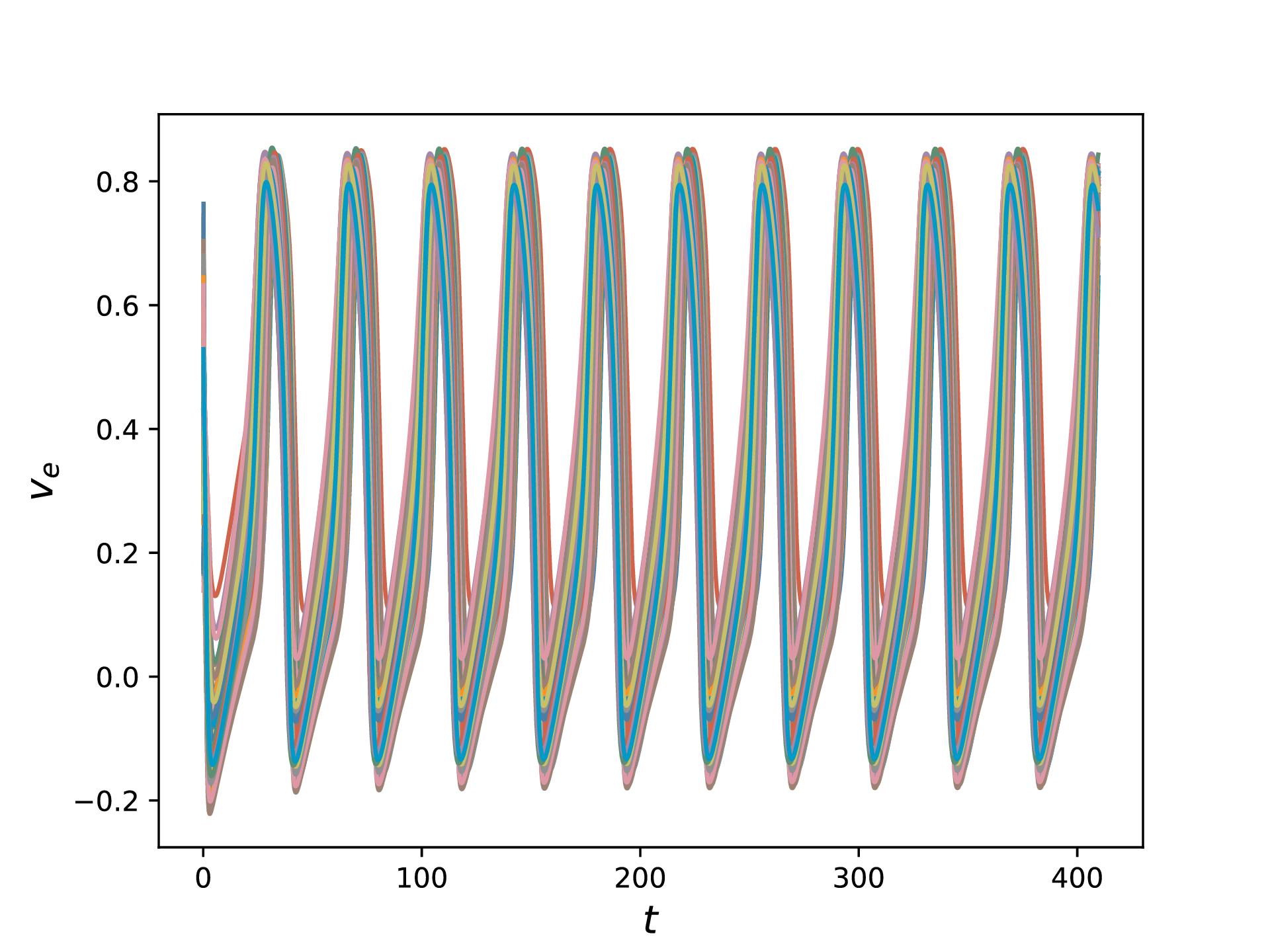}\hfill
        \includegraphics[width=0.5\linewidth]{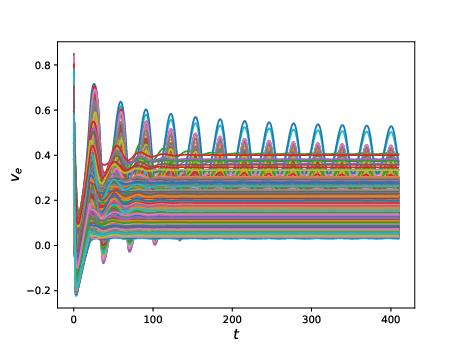}\hfill
        \includegraphics[width=0.5\linewidth]{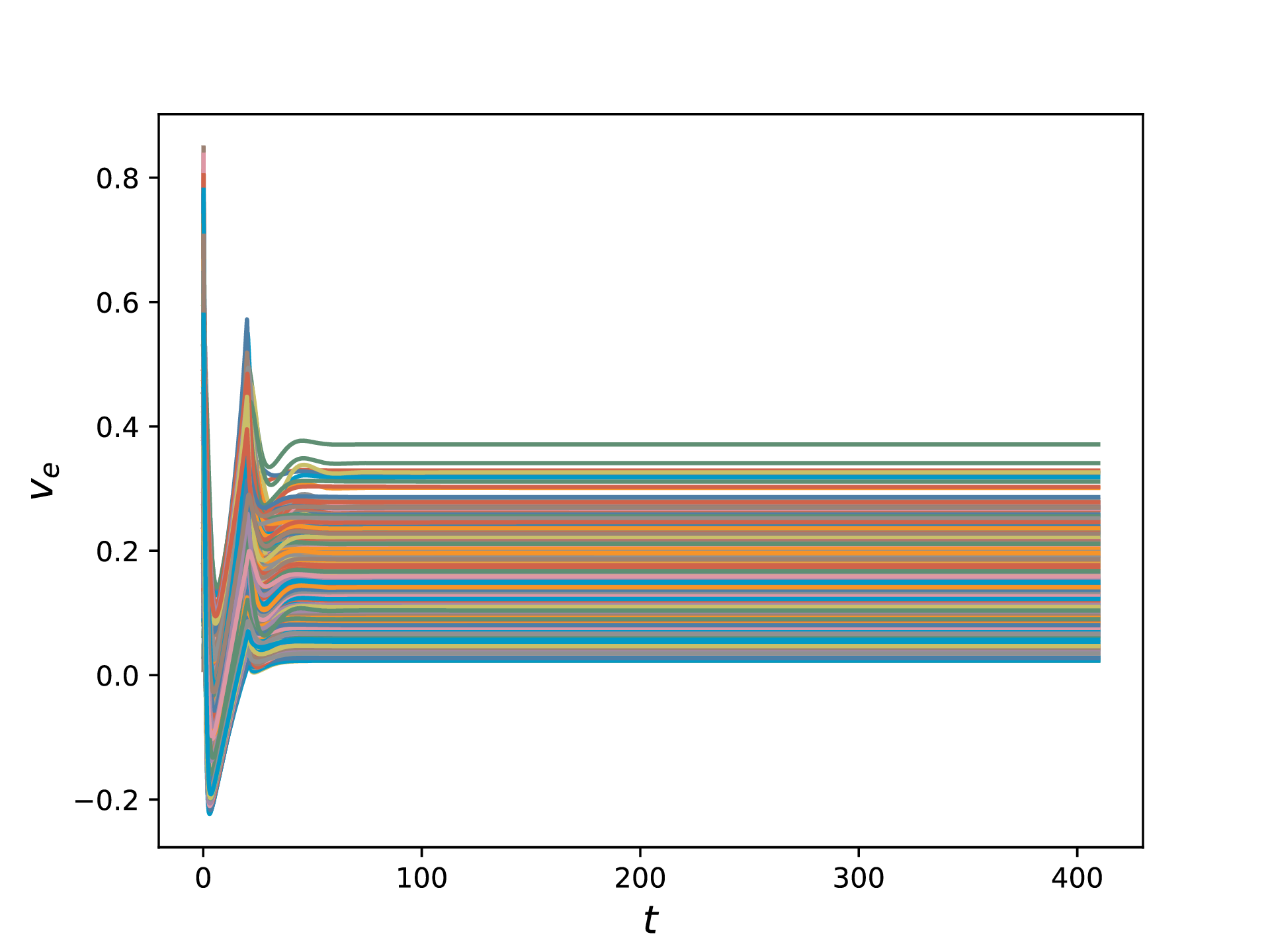}
      \end{minipage}
    }
  \end{minipage}
  \hfill
  \begin{minipage}[t]{0.49\textwidth}
    \centering
    \subcaptionbox{\label{figDS:newfig}}[
      \linewidth]{\includegraphics[width=\linewidth]{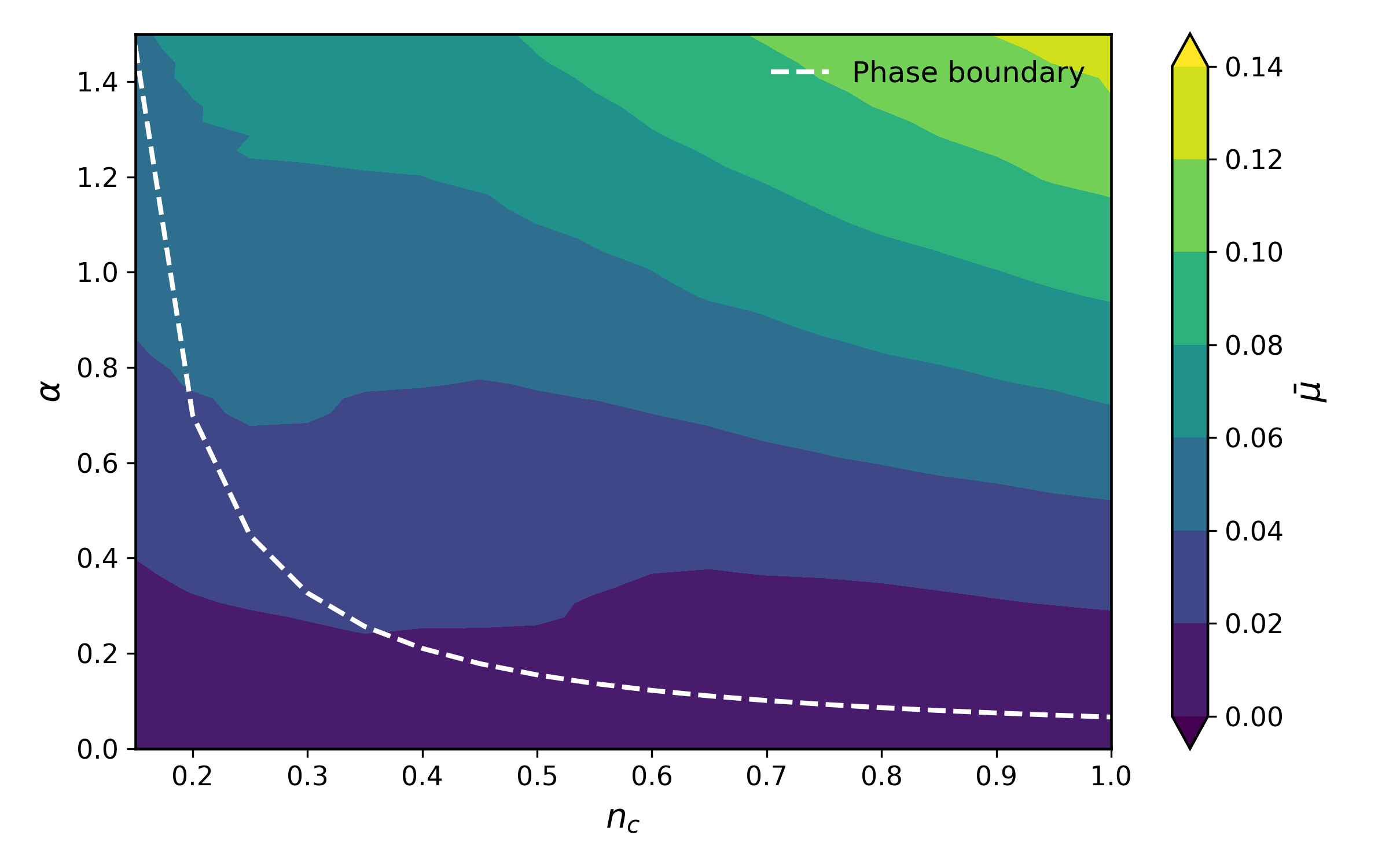}}\par\vspace{0.8em}
    
    \subcaptionbox{\label{figDS:fig4}}[
      \linewidth]{\includegraphics[width=\linewidth]{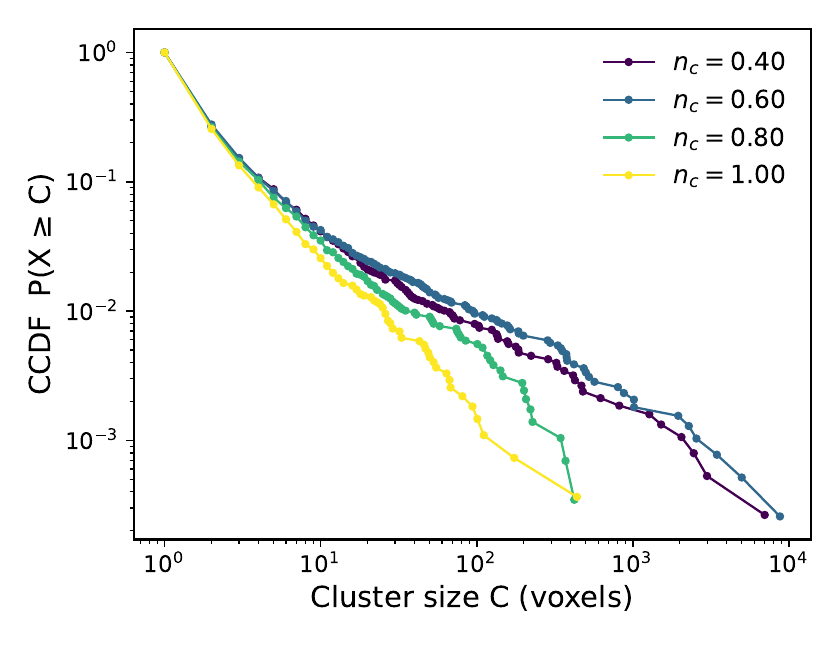}}
  \end{minipage}

  \caption{Sparse control in the global–synchronization regime.  
    \textbf{Left column (deterministic).}  
    (a) Phase boundary separating controlled ($f_o=0$) and non-controlled ($f_o>0$) phase, as a function of $n_c$ and $D$;  
    (b) Temporal dependence of the membrane potentials for three representative cases: uncontrolled (top left, $f_o=1$), `almost' controlled (top right, $f_o \simeq 0$), and fully controlled (bottom, $f_o = 0$).  
    \textbf{Right column (stochastic).}  
    (c) Filled contours of control effort in the $n_c{-}\alpha$ plane for $D=1$ and noise amplitude $\sigma= 0.05$.
        Superimposed on the heatmap is a white dashed curve indicating the phase boundary $\alpha_c(n_c; D=1)$, i.e., the set of $(n_c,\alpha)$ values that separates the fully controlled regime ($f_o=0$) from the oscillatory regime ($f_o>0$) as determined in Fig.~\ref{figDS:main}.
    (d) CCDF of the spatial size of spatio-temporal clusters of oscillating nodes (``wave clusters'') for $\alpha=\alpha_c(n_c)+0.01$ for different values of $n_c$ and noise amplitude $\sigma= 0.05$.}
  \label{fig:distributedsystem}
\end{figure}

Moving to the spatially extended system described in Eqs. \ref{eq:sys} but in the absence of noise, we consider parameters such that the system operates in the  \emph{global synchronization} regime. This means that, in the absence of control, all elements in the medium would oscillate at the same frequency, with a single wave traversing the entire system (see Fig. \ref{figDS:examples} top-left). Let us define $f_o$ as the fraction of oscillating sites. A perfectly working control is expected to suppress all oscillations, leading to $f_o=0$. Defining $p_{i}$ as the power of the mean-removed signal $V_i$, $p_{th}$ as a threshold value below which the activity is deemed negligible and $H(x)$ being the heavy-side function,
$f_o$ characterizes the control performance:
\begin{equation}
\label{eq:metrics_control}
 f_o=\frac{1}{L^2}\sum_{i}H(p_{i}-p_{th})~.
 \end{equation}
The control is achieved via efforts quantified by the long-time power exerted by the control system:
 \begin{equation}
 \bar{\mu}=\frac{1}{L^2} \sum_{i} \Bigg(\frac{1}{T-T_{eq}} \int_{T-T_{eq}}^{T} dt~ \mu^2_{i}(t)\Bigg).
 \label{thwgrbq}
\end{equation}
Recall that, as stated in Eqs. \eqref{eq:sys}, for sites $i$ that are not among the controlled ones, $\mu_{i}(t)=0 \enspace \forall t$.

For each pair $(D,n_c)$ we estimate the critical gain $\alpha_c$, defined as the $\alpha$ at which the system transitions from a fully controlled state ($f_o=0$) to a non-controlled one with $f_o>0$\footnote{We verified robustness to the choice of $p_{th}$ over several orders of magnitude.}. The critical gain $\alpha_c$ is determined from an ensemble of $M=10$ independent realizations. In each realization $m=1,\dots,M$, (i) we draw the set of controlled sites $C$ uniformly at random with its cardinal fixed at the value $|C|=n_cL^2$; (ii) we sample the numbers $\{n_{p,i}\}$ of passive cells for each active cell as i.i.d. random variables taken from a Poisson distribution with mean $f$; (iii) we initialize $(V,g,V_p,\mu)$ at small random values. The dynamics is then \emph{deterministic} (noiseless). 

We integrate Eqs.~\eqref{eq:sys} with a fixed time step for a total of $T$ steps, discarding the first $T_{\rm eq}$ steps (equilibration). Over $[T_{\rm eq},T]$, we compute $p_i$ as the mean square of $V_i$ after removing its time average on that same window. We scan $\alpha$ on a grid (bisection method); for each $\alpha$, we record $f_o(\alpha)$ at the end of the window. The realization–level critical gain $\alpha_c^{(m)}$ is defined as the smallest $\alpha$ for which $f_o(\alpha)=0$. 

In Fig.~\ref{figDS:main}, the curves show the mean $\widehat{\alpha}_c(D,n_c)$  and the shaded bands show  the standard deviation $\mathrm{SD}[\alpha_c]$ over the $M=10$ realizations. Curves terminate at the smallest $n_c$ for which some $\alpha$ can suppress oscillations. 

Obviously, for $D=0$, every excitable site must be directly controlled ($n_c=1$) to suppress oscillations. For $D>0$, diffusive coupling allows the effect of control to spread, so smaller $n_c$ can still stabilize the lattice. As the control gain $\alpha$ is decreased from large values, the fraction of oscillating sites $f_o$, which here plays the role of an order parameter, remains exactly zero above a well-defined threshold $\alpha_c(D,n_c)$ and then increases smoothly from zero as $\alpha$ falls below $\alpha_c$ (Figs.~\ref{figDS:examples} top-right, \ref{fig:foscvsalpha}). This smooth onset of activity corresponds, in the language of bifurcation theory, to a supercritical (continuous) transition in $f_o$ as a function of the control parameter $\alpha$, in contrast to a subcritical (first-order) transition where $f_o$ would jump discontinuously at $\alpha_c$.

At low control coverage (low $n_c$), we observe a large spread of the critical control gain  $\alpha_c$ across realizations. Our working hypothesis, substantiated in Section \ref{sec:si_onset_variability}), is that the onset is governed by a largest-cluster effect, which can be formulated within extreme value theory: global oscillations can be maintained provided that a sufficiently large uncontrolled cluster $S_{\max}$ of connected cells continues to self-sustain, even in the presence of leakage from diffusive control. Realizations with larger, more weakly connected $S_{\max}$ (e.g.\ greater geodesic diameter or smaller algebraic connectivity $\lambda_2$) are harder to extinguish and thus exhibit a \emph{higher} critical control threshold value $\alpha_c$. As the fraction $n_c$ of controlled cells increases, $S_{\max}$ shrinks and becomes highly exposed to control, so these extreme value statistical effects fade and the large variability from realisation to realisation collapses.

In contrast, the inflection point $\alpha_{50}$, the $\alpha$ at which roughly half of the lattice oscillate, is a \emph{system-wide}, self-averaging property: its coefficient of variation is small across the set of different realisations and its value is well predicted by the Hopf threshold of a simple two-cell heterogeneous mean-field model (see Section~\ref{sec:bulk}).


To quantify the intensity of control required to stabilize the system, we analyze the behavior of $\bar{\mu}$ [Eq.~\eqref{thwgrbq}] in the presence of noise. Figure~\ref{figDS:newfig} presents a heatmap of the control effort in the $(n_c,\alpha)$ plane for $D=1$, a regime which, in the absence of control, leads to a globally synchronized oscillatory state. Superimposed on the heatmap is a white dashed curve indicating the phase boundary $\alpha_c(n_c; D=1)$, i.e., the set of $(n_c,\alpha)$ values that separates the fully controlled regime ($f_o=0$) from the oscillatory regime ($f_o>0$) as determined in Fig.~\ref{figDS:main}. As expected, the control effort increases with both $\alpha$ and $n_c$. Note that a ``strategy'' with large $n_c$ and low $\alpha$ is less costly than one with small $n_c$ and large $\alpha$. 

The fact that a control strategy with low $n_c$ and high $\alpha$ is more costly than the high $n_c$-low $\alpha$ counterpart can be understood by examining the distribution of spatio–temporal traveling–wave sizes in the vicinity of the \emph{phase boundary} $\alpha_c(n_c; D)$, defined as the curve in the $(n_c,\alpha)$ plane separating the fully controlled regime ($f_o=0$) from the oscillatory regime ($f_o>0$) (cf. Fig.~\ref{figDS:main}). For $D=1$, this boundary plays the role of a ``critical line'' in the control–parameter space: close to it, the system is marginally stable and exhibits large, noise–driven wave clusters whose statistics depend strongly on $n_c$, thereby modulating the required control effort.

The exact procedure for computing the distribution of spatio–temporal traveling–wave sizes is adapted from Ref.~\cite{sakaguchi2024noise} and summarised here. We first compute a binary activity field $b_i(t)$ that equals 1 if the instantaneous amplitude of site $i$ exceeds a small threshold \emph{and} all its nearest neighbours are also above threshold at the same time step; otherwise $b_i(t)=0$. This local–neighbour condition suppresses spurious single–site activations due to noise. We then identify connected components in the $(x,y,t)$ space–time volume by grouping nearest–neighbour active sites in both space and time. Each connected component is defined as a \emph{wave cluster}, and its size $C$ is given by the total number of active site–time points it contains. The distribution $P(C)$ is obtained by collectif the set of clusters generated over a long time interval.

Figure~\ref{figDS:fig4} shows the complementary cumulative distribution function (CCDF) of the wave–cluster size $C$ for different $n_c$, with $\alpha$ set just above the phase boundary, $\alpha = \alpha_c(n_c; D=1) + 0.01$. The tails of these CCDFs are approximately power–law, with heavier tails for smaller $n_c$. This indicates that, near the phase boundary, systems with fewer controlled sites experience larger noise–driven wave clusters, which in turn require greater average control effort $\bar{\mu}$ to maintain stability compared to systems with larger $n_c$.

Ideally, the control should strive for a trade-off between robustness and flexibility, while minimizing its energy consumption: on the one hand, external perturbations of the parameters should not cause abrupt and undesired change of state, but on the other hand the body has to keep the ability to efficiently switch function. Depending on the exact characterization of this trade-off, the optimal strategy is in the vicinity of the phase transition\cite{magnasco2022robustness}.

\subsection{Dynamical Bifurcations}
This section introduces a new important element in the form of a slowly varying control parameter. The added realism of these two factors (noise and varying control parameter) is important to test our main hypothesis: pre-labor contractions are crucial from the control perspective to maintain the right balance between quiescence and contractility.

In the preceding section, we analyzed a \emph{static} bifurcation, first at the level of a single cell and then for the entire network. This was done by examining the system after it had reached equilibrium from randomly chosen initial conditions and under fixed parameter values (e.g., $\alpha$). In contrast, a \emph{dynamical} bifurcation \cite{kuehn2015multiple} involves a qualitative change in the system’s state induced by slowly varying a parameter across its critical threshold. Since we envision control approaching the critical point to minimize cost, we now treat $\alpha$ as a time-dependent parameter:
\begin{equation}
\label{eq:slow_alpha}
    \alpha(t)=\alpha_0-rt, \quad 0<r\ll 1.
\end{equation}

\begin{figure}[H]
  \centering
  \subfloat[\label{figOU:fig1}]{\includegraphics[width=0.5\textwidth]{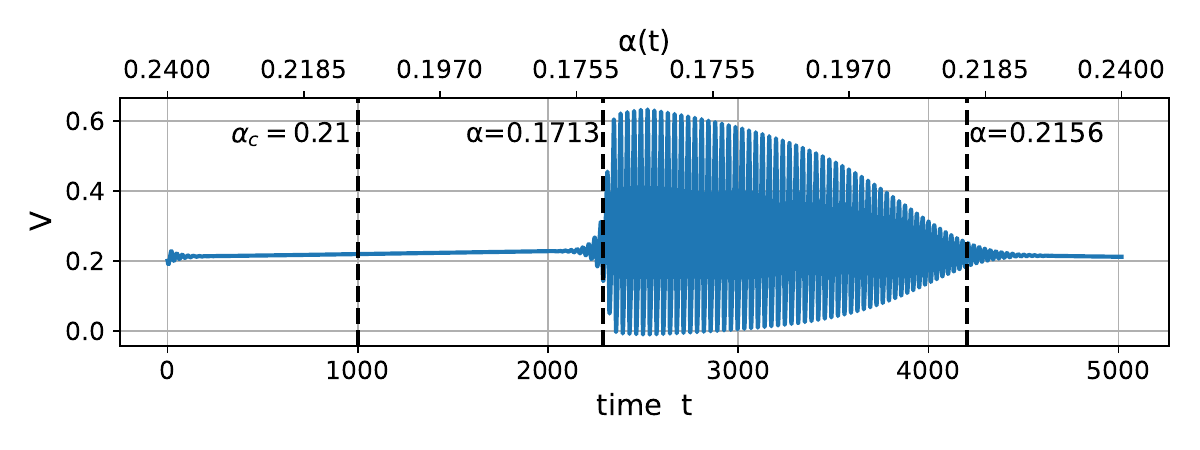}}
  \hfill
  \subfloat[\label{figOU:fig2}]{\includegraphics[width=0.5\textwidth]{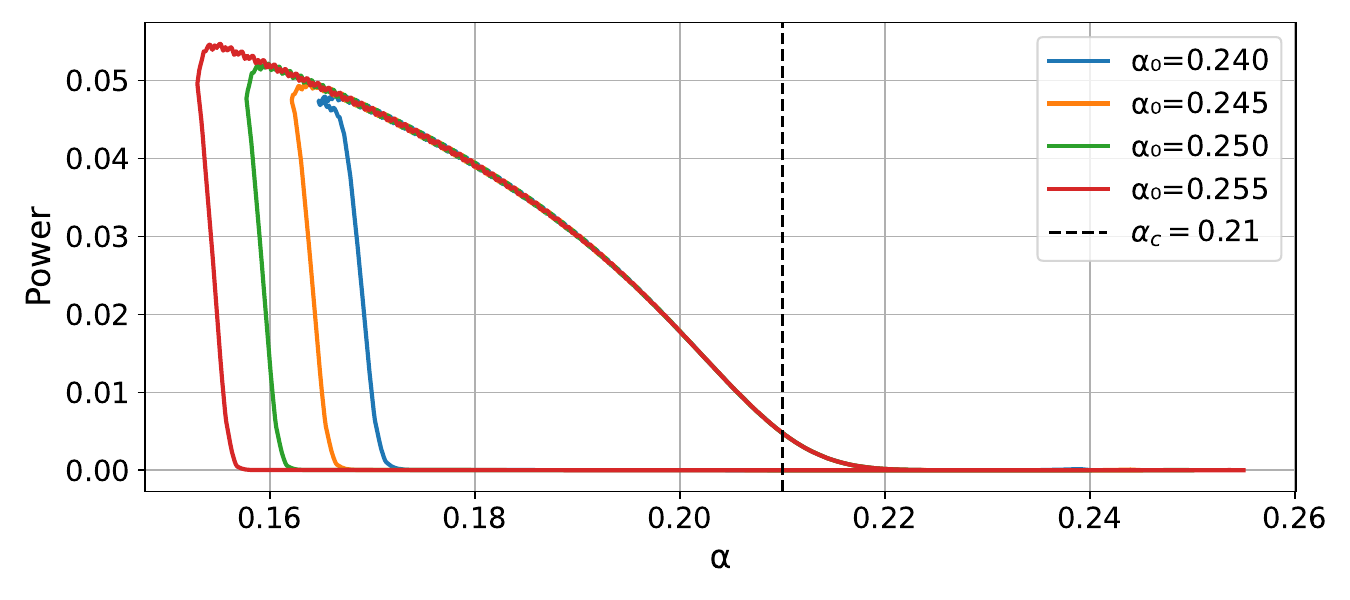}}

  \vspace{1em}

  \subfloat[\label{figOU:fig3}]{\includegraphics[width=0.45\textwidth]{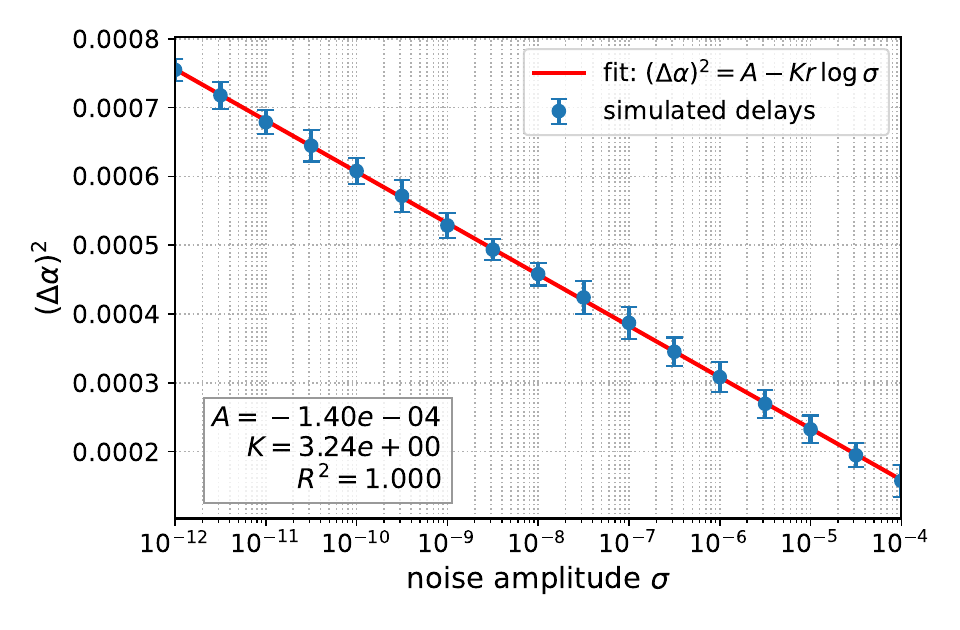}}
  \hfill
  \subfloat[\label{figOU:fig4}]{%
    \raisebox{0.3cm}{\includegraphics[width=0.5\textwidth]{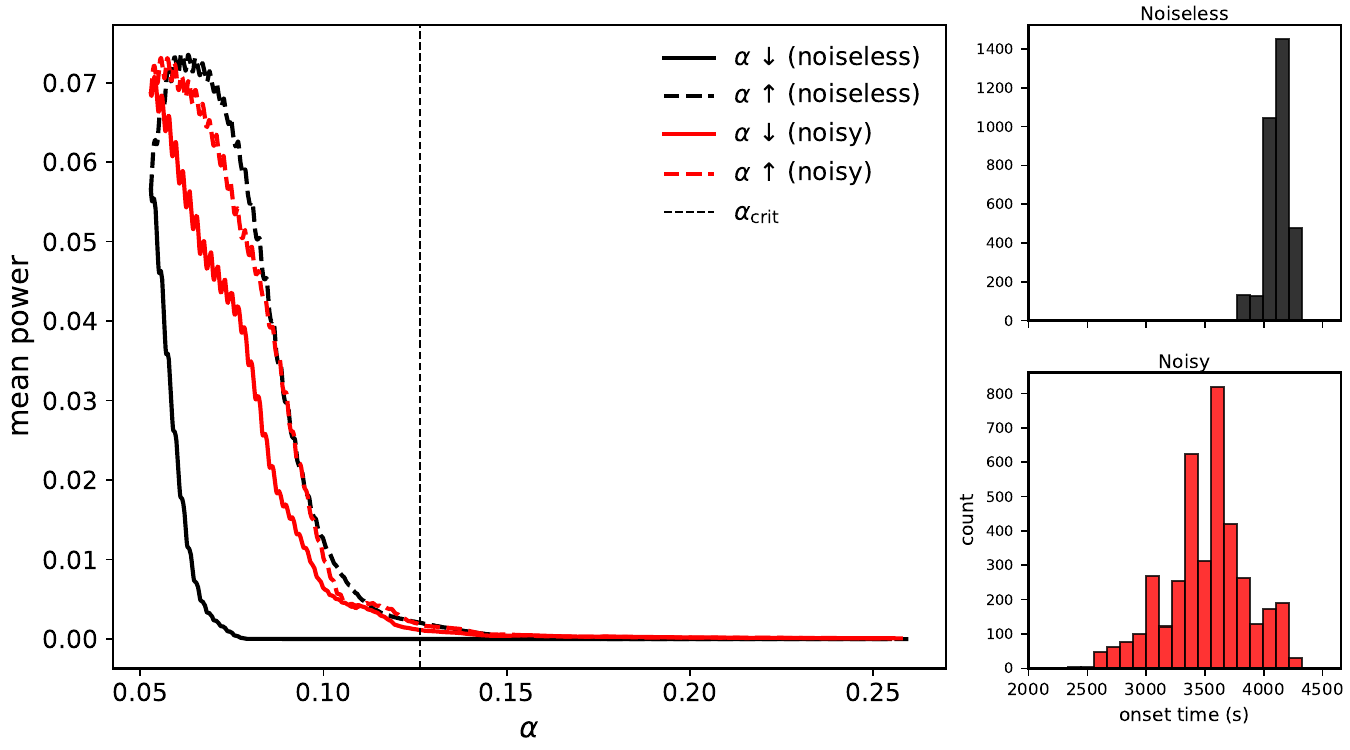}}}

  \caption{Dynamic bifurcations. 
(a) Voltage over time for one isolated cell (Eq.~\ref{eq:sys1dPRIMA}) with slowly varying $\alpha$. The dashed lines indicate, respectively from left to right, the static bifurcation point, the actual onset of oscillation, and the point at which oscillations are suppressed again. Note how the first and third $\alpha$ values are virtually identical, implying that the Hopf bifurcation delay is present only in one direction. 
(b) Power of the electric potential $V$ over a sliding window as a function of $\alpha$, first slowly decreasing and then increasing back, for different starting points $\alpha_0$. Parameters for (a) and (b) are the same ($r=10^{-5}$, $C_r=1$, $n_p=1$). (c)\; Square of the mean bifurcation shift, $(\Delta\alpha)^2$, versus the noise amplitude $\sigma$ (standard deviation of the additive noise) on a logarithmic $x$-axis for fixed $\alpha_0$.
Here $\Delta\alpha = \langle \alpha_{\mathrm{onset}} \rangle - \alpha_c$ is averaged over 100 independent realizations (error bars show standard deviations). 
(d) Hysteresis in the spatially extended system. Parameters: $n_c=0.7$, $D=1$, $r=10^{-5}$. The histograms show the onset–oscillation time on the $\alpha$-decreasing branch with and without noise.} 
  \label{figOU:four_grid}
\end{figure}

Let us start from a single cell in absence of noise (Eq. \ref{eq:sys1dPRIMA}). Fig \ref{figOU:fig1} shows that the transition occurs only when the parameter has moved well past the value predicted by standard bifurcation analysis, that assumes a static, fixed control parameter and ignores its dynamic evolution.  Such delayed Hopf bifurcation, also called \emph{slow passage through a Hopf
bifurcation}, is a well known phenomenon \cite{baer1989slow, neishtadt1987persistence}. The intuition behind this effect is the following. Near the equilibrium, the linearization has a complex eigenpair
$\lambda(\alpha)=q(\alpha)\pm i\omega(\alpha)$ with $q(\alpha_c)=0$ ($\alpha_c$ is the critical value predicted by the static bifurcation analysis).
As the control drifts according to $\dot\alpha=-r$ with $0<r\ll1$, the mode amplitude
$A(t)$ obeys $\dot A \approx q(\alpha(t))\,A$, so growth is governed by the
\emph{accumulated} positive rate $\int q(\alpha(s))\,ds$.
Although linear stability is lost when $\alpha$ crosses $\alpha_c$,
the instantaneous growth rate $q(\alpha(t))$ is positive but initially small, so
the trajectory
continues to shadow the now-unstable equilibrium until the \emph{accumulated} positive
growth balances the earlier negative growth. In terms of the real part \(q\), the onset
time \(t_{\mathrm{onset}}\) (the time at which sustained oscillations are first observed in the dynamic sweep) is well approximated by the integral balance
\begin{equation}
\label{eq:int_balance}
\int_{t_0}^{t_{\mathrm{onset}}} q\big(\alpha(s)\big)\,ds \;\approx\; 0
\quad\Longleftrightarrow\quad
\frac{1}{r}\!\int_{\alpha_{\mathrm{onset}}}^{\alpha_c}\! q(\alpha)\,d\alpha
\;=\;
\frac{1}{r}\!\int_{\alpha_c}^{\alpha_0}\! \big|q(\alpha)\big|\,d\alpha,
\end{equation}
where $\alpha_{\mathrm{onset}}$ is the value of $\alpha$ at which sustained oscillations are first observed in the dynamic sweep. Equation \ref{eq:int_balance} implies \(\alpha_{\mathrm{onset}}<\alpha_c\). This is due to the fact that, for $\alpha > \alpha_c$, $q(\alpha)$ is negative. Hence for the integral over $q\big(\alpha(s)\big)$ to be zero,  $\alpha_{\rm onset}$ should be smaller than $\alpha_c$ so that the positive values of $q(\alpha)$ for $\alpha_{\mathrm{onset}}<\alpha < \alpha_c$ compensate the negative values of $q(\alpha)$ for $\alpha_c <  \alpha < \alpha_0$. This  explains why the shift increases when
the start \(\alpha_0\) lies farther above from \(\alpha_c\). For a canonical FiztHugh-Nagumo model, in the limit $r\to 0$, Baer et al. \cite{baer1989slow} show that $\alpha_{\mathrm{onset}}$ obeys $\alpha_c-\alpha_{\mathrm{onset}}=\alpha_0-\alpha_c$ independent of $r$, which we verified numerically.

Due to the time-dependent evolution of the control parameter, such system can exhibit hysteresis. Fig. \ref{figOU:fig2} displays the energy of the electric potential $V$ under a time evolving $\alpha(t)$ that initially decreases and subsequently increases. Each trajectory has a different starting point $\alpha_0$. Hysteresis is one mechanism that can give rise to catastrophic transitions \cite{scheffer2001catastrophic}. In our case, consider a scenario where the control gradually decreases $\alpha(t)$
over time while monitoring the onset of oscillations. Due to hysteresis, once oscillations emerge, it may already be too late for the control to restore stability—potentially resulting in excessively strong contractions or even triggering parturition.

This is where noise plays a beneficial role. Consider applying an additive white noise $\chi_v(t) \neq 0$ of standard deviation $\sigma$ to the first equation of the potential $V$ in the set of Eq.~\ref{eq:sys1dPRIMA} while slowly decreasing $\alpha$ according to Eq.~\ref{eq:slow_alpha}.
 To quantify the effect of noise, we define the \emph{mean bifurcation shift} 
\begin{equation}
\Delta\alpha = \langle \alpha_{\mathrm{onset}} \rangle - \alpha_c, 
\end{equation}
where once again $\alpha_{\mathrm{onset}}$ is, for each realization, the value of $\alpha$ at which sustained oscillations are first observed in the dynamic sweep, and $\langle \cdot \rangle$ denotes the average over independent realizations. The reference $\alpha_c$ is the critical value predicted by the static bifurcation analysis (Sec.~\ref{secsub:scellcontrol}).

In the absence of noise, $\Delta\alpha < 0$ due to the slow–passage effect described above (see Eq. \ref{eq:int_balance}).
Figure~\ref{figOU:fig3} plots $(\Delta\alpha)^2$ versus the noise amplitude $\sigma$
on a logarithmic $x$-axis for a fixed $\alpha_0$). We observe that $(\Delta\alpha)^2$ decreases linearly with $\log \sigma$, in agreement with theoretical predictions for the stochastic Hopf normal form \cite{berglund2006noise}.
Physically, noise perturbs the system out of the slow–passage bottleneck sooner, shortening the deterministic delay and enabling oscillations to commence nearer to $\alpha_c$. In this way,  stochastic fluctuations reduce hysteresis by narrowing the parameter range where the system rests in a metastable state.

We now extend our analysis to the 2D lattice. Fixing $n_c$ and selecting a point to the right of the transition curve in Fig.\ref{figDS:main}, we slowly decrease $\alpha$ past the transition and then increase it back. This procedure is repeated both in the presence and absence of noise applied to the set of potentials $\{V_i\}$ (see Eqs.\ref{eq:sys} and \ref{eq:slow_alpha}). Figure~\ref{figOU:fig4} displays the mean power per cell, computed over a sliding window, as a function of $\alpha$, along with a histogram of oscillation onset times during the decreasing phase of $\alpha$. The 2D lattice exhibits qualitatively similar bifurcation delay behavior, consistent with observations in networks of FitzHugh–Nagumo oscillators \cite{premraj2018bifurcation}. Notably, the introduction of noise eliminates the delay, indicating that fluctuations in uterine contractions can enhance control by preventing undesirable hysteresis.
This finding highlights a key insight: fluctuations in uterine contractions can play a beneficial role in control by mitigating bifurcation delay.

\subsection{Early warning signals of parturition}
\label{sec:ews}
\begin{figure}[]
  \centering
  \includegraphics[width=0.6\linewidth]{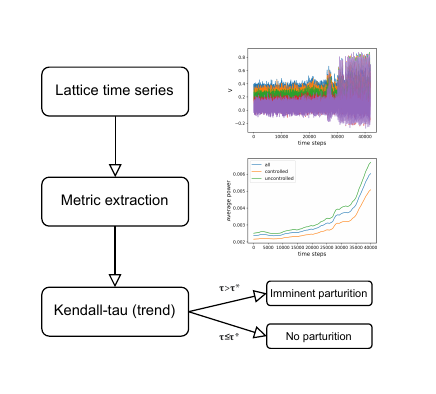}

  \caption{%
    Classification pipeline for detecting imminent parturition. The monitoring module processes the time series of membrane potentials from individual cells to extract a one-dimensional summary metric. It then computes the trend using the Kendall tau coefficient. If the trend exceeds a predefined threshold, the instance is classified as indicating imminent parturition.
  }
  \label{fig:pipeline}
\end{figure}

Thus far, we have considered a control strategy that, aiming to stabilize the system while minimizing cost, gradually reduces its feedback strength $\alpha$ toward the critical line for a fixed fraction of controlled cells $n_c$.
We now pose the following question:

\begin{center}
Can the control accurately detect its operational limits?
\end{center}

We propose an approach that is conceptually related to the broader literature on early warning signals (EWS), which typically rely on critical slowing down or increasing variance of fluctuations to anticipate bifurcations. As a system nears a critical point, its recovery from perturbations slows down, leading to increased variance, rising spatial and temporal correlations, and related statistical signatures \cite{scheffer2001catastrophic, carpenter2006rising, bury2020detecting}.

Specifically, we introduce the concept of an internal sentinel module that continuously monitors a simple early-warning statistic and triggers a corrective ``hormonal pulse'' when signs of impending loss of quiescence are detected. In this paper, we implement and evaluate the \emph{detection} module only; how such pulse would couple to the model is outlined for future work in Sec.~\ref{sec:discussion}. 
This assumption aligns with the broader framework of predictive homeostasis \cite{simor2023sleep, deans2021biological}, in which physiological control systems exploit early-warning fluctuations to maintain stability. It supports the idea that the endocrine–myometrial network may similarly use such cues to preserve uterine quiescence.

Figure~\ref{fig:pipeline} presents a flowchart of our proposed sentinel module for early detection of critical transitions. Starting from the multivariate time series of membrane potentials, the module computes a one-dimensional summary statistic, which is the average power per cell $\overline{P}_{\mathcal{S}}(t)$ over a sliding time window. To define $\overline{P}_{\mathcal{S}}(t)$, we proceed as follows. In line with our previous analysis, we adopt the average power per cell as a simple yet informative early-warning metric. Let $\Tilde{V}_i$ denote the mean-removed voltage signal of cell $i$ over a time window of length $T$; the corresponding signal power is
\begin{equation}
    p_i(t)=\frac{1}{T}\int_{t}^{t+T} \Tilde{V}_i^2(s)\,ds~.
\end{equation}
For a given subset of sites  
\(\mathcal{S}\in\{\mathcal{C}\;(\text{controlled}),\;
               \mathcal{U}\;(\text{uncontrolled}),\;
               \mathcal{L}\;(\text{all})\}\),
we define the average power per cell as
\begin{equation}
\label{eq:avg_pow_cell}
\overline{P}_{\mathcal{S}}(t)=
\frac{1}{|\mathcal{S}|}\,
\sum_{i\in\mathcal{S}} p_i(t)~.
\end{equation}
For example, with $n_c=0.6$ and $\mathcal{S}=\mathcal{U}$, the sentinel monitors only the $40\%$ of cells that are uncontrolled.

To assess whether $\overline{P}_{\mathcal{S}}(t)$ is increasing or decreasing, we calculate the \emph{Kendall rank correlation coefficient} $\tau_K$ between $\overline{P}_{\mathcal{S}}(t)$ and time $t$ over the most recent $W$ data points:
\begin{equation}
    \tau_K = \frac{\text{number of concordant pairs} - \text{number of discordant pairs}}{\binom{W}{2}}.
\end{equation}
This nonparametric statistic ranges from $-1$ (strictly decreasing trend) to $+1$ (strictly increasing trend), with $\tau_K \approx 0$ indicating no monotonic trend. A significantly positive $\tau_K$ indicates that the monitored metric is trending upward—consistent with an impending loss of quiescence—and triggers the release of a corrective hormonal pulse. If $\tau_K$ exceeds a predefined classification threshold, the system is flagged as approaching parturition, prompting the control module to halt the ongoing decrease of $\alpha$. This approach is robust to non-Gaussian fluctuations and does not require assuming a specific functional form for the trend.
 
In order to \emph{evaluate} the ability of the sentinel module to anticipate parturition-like transitions, we treat it as a binary classifier operating on time–series data. For a given pair \((n_c,D)\) and a given noise standard deviation $\sigma$, we generated \(N=100\) independent multivariate time series in two distinct regimes.  

In the \emph{transition} regime, the control parameter \(\alpha(t)\) starts at a value such that the system is under control, and is then ramped down slowly towards the phase boundary, eventually triggering a loss of quiescence.  

In the \emph{no-transition} regime, \(\alpha(t)\) follows an equally slow descent but remains at all times a fixed margin above the critical value, representing conditions under which the controller can safely continue minimizing its cost without approaching instability.  

These two ensembles serve as, respectively, the positive (“approaching parturition”) and negative (“remaining quiescent”) classes for the classifier (see Sec.~\ref{sec_app:EWS} for details).  

By sweeping a classification threshold $\theta$ on $\tau_K$, we obtain the \emph{true positive rate} (TPR) and \emph{false positive rate} (FPR):
\[
\mathrm{TPR}(\theta) = \frac{\text{\# positives with }\tau_K > \theta}{\text{\# positives}}, \quad
\mathrm{FPR}(\theta) = \frac{\text{\# negatives with }\tau_K > \theta}{\text{\# negatives}}.
\]
In biological terms:
\begin{itemize}
    \item \emph{True positive}: an imminent parturition correctly detected by the sentinel module.  
    \item \emph{False positive}: a quiescent state incorrectly flagged as approaching parturition — potentially wasting energy and resources. 
    \item \emph{True negative}: a quiescent state correctly recognized as safe. 
    \item \emph{False negative}: an imminent parturition that goes undetected — the controller fails to react in time.    
\end{itemize}

The \emph{receiver operating characteristic} (ROC) curve plots TPR versus FPR as $\theta$ is varied, and the \emph{area under the ROC curve} (AUC) summarizes this trade-off into a single scalar: $\mathrm{AUC}=1$ corresponds to perfect separability, $\mathrm{AUC}=0.5$ to random guessing.  

\begin{figure}[htbp]
  \centering

  \subfloat[\label{fig:auc}]{%
    \includegraphics[width=0.48\textwidth]{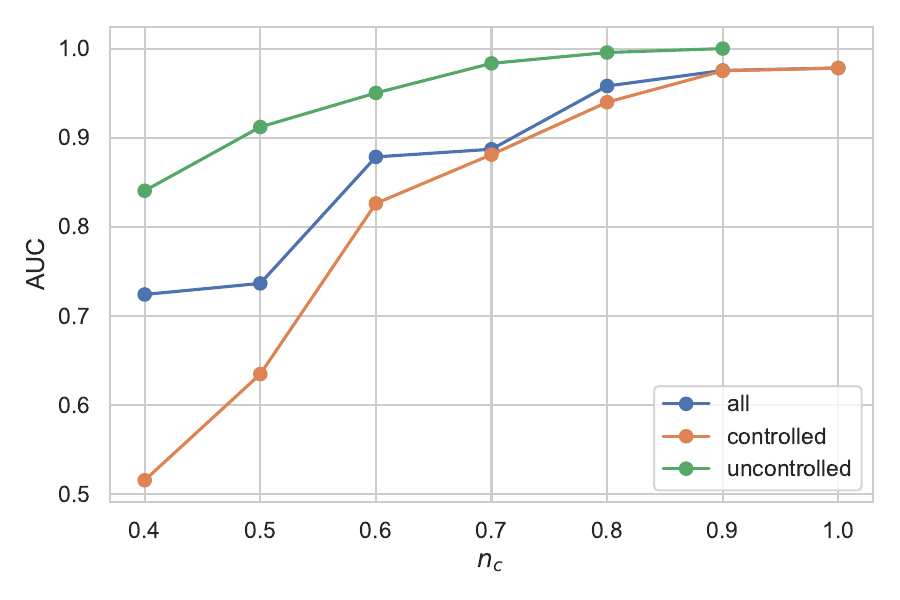}}
  \hfill
  \subfloat[\label{fig:fpr-at-tpr}]{%
    \includegraphics[width=0.48\textwidth]{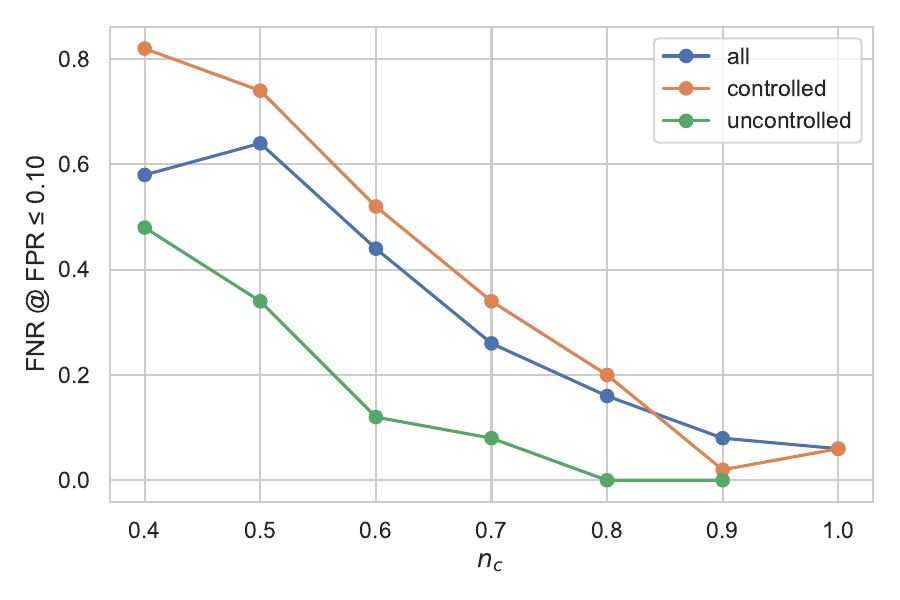}}

  \vspace{1.0em} 

  \subfloat[\label{fig:tau_box_cont}]{%
    \includegraphics[width=0.48\textwidth]{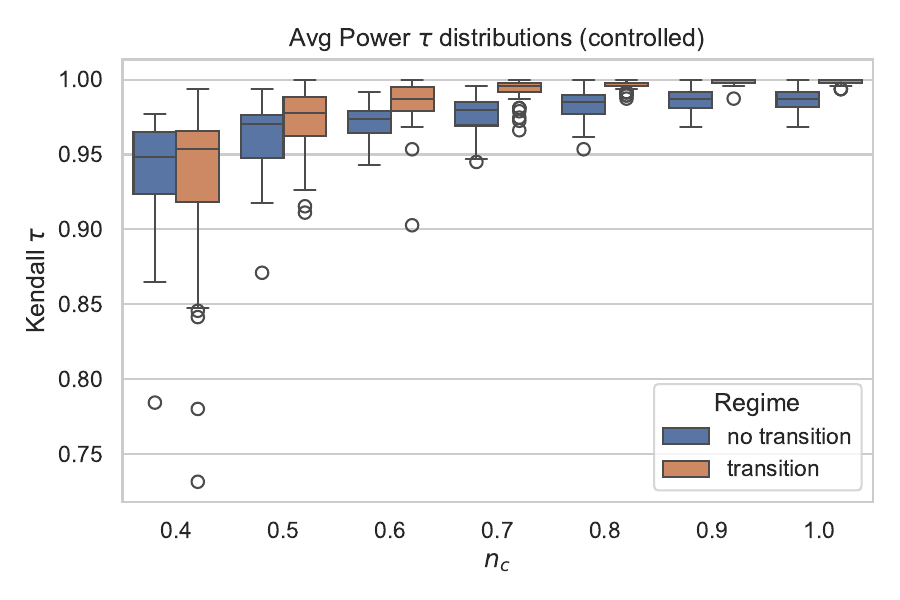}}
  \hfill
  \subfloat[\label{fig:tau_box_uncont}]{%
    \includegraphics[width=0.48\textwidth]{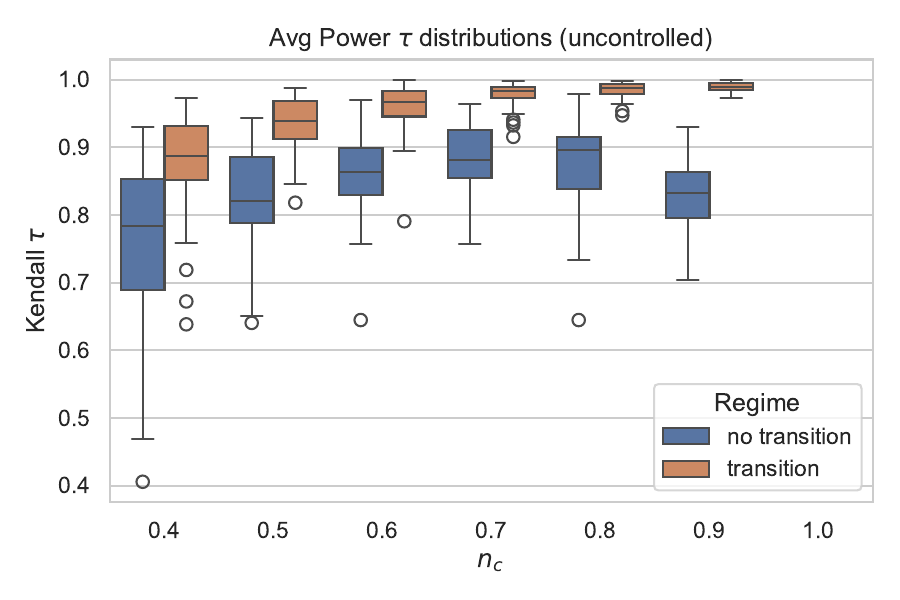}}

  \caption{Performance of the sentinel module treated as a binary classifier.  
(a) Area under the ROC curve (AUC) as a function of $n_c$ for $D=1$.  
Each curve corresponds to a different subset of sites observed by the module: controlled cells, uncontrolled cells, or all cells. The AUC summarizes the trade-off between true positive rate and false positive rate when the decision threshold on $\tau_K$ is varied.  
(b) False negative rate (FNR) when the false positive rate (FPR) is constrained to be below 10\%.  
Here, $\mathrm{FNR} = \frac{\# \text{ imminent parturitions not detected}}{\# \text{ imminent parturitions}}$ and  
$\mathrm{FPR} = \frac{\# \text{ quiescent cases incorrectly flagged}}{\# \text{ quiescent cases}}$.  
The notation “FNR @ FPR” means the FNR measured under the constraint that FPR $< 0.1$.  
(c–d) Box plots of the \emph{Kendall rank correlation coefficient} $\tau_K$ between the average power per cell $\overline{P}_{\mathcal{S}}(t)$ (Eq.~\ref{eq:avg_pow_cell}) and time $t$, computed over the most recent $W$ points of each time series.  
Panel (c) shows the box plots of $\tau_K$ when $\mathcal{S}$ is the set of controlled cells; panel (d) shows box plots of $\tau_K$ when $\mathcal{S}$ is the set of uncontrolled cells.  
In each panel, box plots are shown separately for the “approaching parturition” (positive) and “quiescent” (negative) classes, illustrating the degree of overlap between the two classes in $\tau_K$ space and thus how easily they can be separated by a threshold.
}
  \label{fig:avgpower-eval}
\end{figure}

Figure~\ref{fig:auc} shows the AUC as a function of $n_c$ with $D=1$ for the three subsets of observed sites. Two conclusions can be drawn:  
(i) Performance increases with increasing $n_c$. This is because the Kendall $\tau_K$ statistic measures the significance of the upward trend in average power as $\alpha$ decreases. When $n_c$ is small, decreasing $\alpha$ affects only a limited subset of cells, so the change in average power is modest. When $n_c$ is large, the same decrease in $\alpha$ perturbs a much larger fraction of the system, producing a stronger, more detectable signal.  
(ii) Uncontrolled cells are more informative than controlled ones: although in the absence of noise they remain quiescent through diffusive coupling, their dynamics can carry early signatures of instability that are easier to pick up, leading to higher classification accuracy. Depending on the chosen metric, this can lead to an interior maximum in the AUC (see Sec.~\ref{sec_app:EWS} and Fig. \ref{fig:maxpeak-eval}).  

Figure~\ref{fig:fpr-at-tpr} complements the AUC analysis by showing the false negative rate — the fraction of imminent parturitions that go undetected — when the false positive rate is constrained to remain below 10\%, a regime motivated by energetic considerations. Finally, Figures~\ref{fig:tau_box_cont} and~\ref{fig:tau_box_uncont} display box plots of $\tau_K$ values for controlled and uncontrolled cells in both classes. These plots visualize class separability in the $\tau_K$ feature: the greater the non-overlap between the two distributions, the easier it is for any threshold-based rule to discriminate between “approaching parturition” and “quiescent” cases. The wider separation observed for uncontrolled cells confirms their higher discriminatory power.

\section{Discussion}
\label{sec:discussion}
We developed an \emph{in silico} framework in which the pregnant uterus is represented as a spatially extended network of FitzHugh–Nagumo elements under sparse adaptive control. This extends the open‑loop picture of \cite{singh2012self}, where increasing the diffusive coupling $D$—representing the formation and strengthening of gap junctions \cite{garfield1995role}—drives the tissue toward global synchrony. Yet human parturition is not governed by a single maturational knob: it emerges from a complex, multi‑timescale biochemical dialogue between fetus, placenta, and mother (endocrine, paracrine, and autocrine loops) that keeps the myometrium quiescent for months and then tips it into coordinated contraction at term \cite{kota2013endocrinology}. This architecture is more naturally read as a \emph{control} problem than as a bare phase transition.

Accordingly, we introduce a minimalist ``control center” that suppresses oscillations during gestation and is deliberately withdrawn at labour onset—an idealization of functional progesterone withdrawal and related endocrine shifts \cite{brown2004mechanisms}. In our formulation, the controller operates near its own stability boundary to minimize energetic cost, echoing the idea that biological systems can \emph{hover} close to bifurcations to reconcile robustness with flexibility \cite{magnasco2022robustness}. While the model is deliberately minimalist, it captures reported regional differences in progesterone sensitivity \cite{ji2024exploring} through the fraction of controlled cells $n_c$. Misidentifying that boundary is one pathological route to premature, system‑wide oscillations; the physiological trigger at term, in contrast, is an active release of the brake.

Finally, the slow maturation of coupling ($D$) and the adaptive tuning of control strength ($\alpha$) are intertwined. For a fixed structural constraint $n_c$, the $\alpha$ needed to maintain quiescence decreases as $D$ grows: early in pregnancy $\alpha$ is high to silence a weakly coupled lattice, but as gap junctions strengthen, the controller can relax while still holding the system below the oscillatory threshold. In our model this trajectory is implemented by cost minimisation plus a sentinel module that monitors fluctuations and nudges $\alpha$ to stay just shy of loss of control.

Within this setting, our model provides a unified resolution to the puzzles outlined in the Introduction by formalizing the robustness–flexibility tradeoff of pregnancy as a control problem. Robustness arises from active feedback that suppresses synchronization across the myometrial network, stabilizing quiescence over long time scales despite intrinsic excitability and noise. Flexibility, however, is not achieved by a gradual softening of this suppression or by drifting toward a bifurcation. Instead, it has a dual origin. First, the controller operates close to its stability boundary, ensuring sensitivity to incipient loss of control. Second—and critically—the underlying uterine dynamics retain a fast intrinsic oscillatory mode that remains latent during gestation. When control is withdrawn, this mode is released on its natural time scale, leading to an abrupt and system-wide transition to coordinated contractions. In contrast to systems such as thermostatic regulation, where removing control produces only slow relaxation, the uterus in our model switches function nearly instantaneously once the inhibitory brake is lifted. This architecture reconciles months-long robustness with the need for rapid, decisive labour onset. By reproducing this robustness–flexibility tradeoff, the model generates concrete, system-level predictions for two longstanding empirical phenomena—pre-labour contractions and preterm birth—which we now discuss in turn.

\paragraph{1. Pre--labour contractions as signatures of a cost--aware controller hovering near criticality.}
Empirical studies applying clustering techniques to EHG recordings \cite{russo2021alvarez, esgalhado2020uterine} have identified two distinct types of pre-labor contractions: Alvarez waves, which are localized, low-amplitude, and high-frequency; and Braxton-Hicks waves, which are more spatially spread, with higher amplitude and lower frequency. Their physiological role has remained debated, often framed as ``muscle training'' for labour \cite{raines2017braxton}. Moreover, the relationship between Alvarez waves, Braxston-Hicks waves and pre-term labor remains unsettled \cite{russo2021alvarez}. In our model, these events emerge naturally when the controller chooses an operating point close (but to the right) of the transition line (Fig.~\ref{figDS:main}) so as to minimise long--term energetic effort while retaining the ability to switch rapidly. The proximity to the critical line together with unavoidable stochastic fluctuations in membrane potential (e.g. channel noise\cite{white2000channel}) imply a broad distribution of spatio--temporal wave sizes, with power law--like tails (Fig. \ref{figDS:fig4}), reflecting a state of dynamical criticality \cite{magnasco2022robustness}. These noise--driven events are therefore not merely ``imperfections'' but informational probes: they sample latent instabilities and feed back to the controller whether its current strength $\alpha$ suffices (see point 3). Current surface‑EHG systems typically use 4×4 grids spaced several centimetres apart; this granularity is sufficient to capture global Braxton‑Hicks bursts but undersamples the finer spatial statistics, making the power law distribution of synchronised cluster sizes predicted by our model difficult to verify directly.

Interpreting these noise–driven contractions as informational probes naturally raises the question of empirical testability. One might attempt to assess whether a pre-labour contraction is followed by systematic changes in subsequent activity—analogous to foreshock/aftershock analyses in seismology—by comparing uterine dynamics before and after such events. However, this type of falsification faces intrinsic limitations. The control system does not respond to contractions per se, but to internal signals that remain unobservable at the tissue level, such as latent excitability, coupling heterogeneity, or endocrine state. Surface EHG recordings capture only the macroscopic electrical consequences of this process, not the variables that the controller itself “sees” or acts upon. As a result, the absence of a detectable post-contraction signature in observed signals does not invalidate a probing interpretation; it merely reflects the fact that multiple control responses—recalibration, confirmation of adequacy, or deliberate inaction—can be consistent with identical observable outcomes. Moreover, alternative mechanisms, such as refractory effects or local recovery dynamics, can generate statistically similar pre/post patterns without involving active information gathering. At the spatial and temporal granularity of current EHG data, these hypotheses are empirically entangled, making definitive discrimination impossible. Resolving this ambiguity would require measurements that directly access finer-scale spatial structure or internal state variables, for example through high-density electrode arrays, regional endocrine readouts, or controlled perturbation experiments.

\paragraph{2. Noise--suppressed hysteresis enhances flexibility.} In absence of noise, a slow decrease of $\alpha$ causes a delayed onset of oscillations, with consequent hysteresis, potentially trapping the uterus in an oscillatory state once contractions appear. Introducing biologically realistic noise eliminates the delay (Fig. \ref{figOU:fig4}), collapsing the forward and backward branches and thereby widening the basin of recoverable quiescence. This mechanism illuminates a paradox: why does the pregnant uterus tolerate (and perhaps generate) low--amplitude fluctuations rather than enforcing a perfectly silent state? Our results suggest that such fluctuations enhance controllability by preventing catastrophic path dependence. Similar noise‑mediated suppression of hysteresis or delay has been reported in slow–fast bifurcations \cite{berglund2006noise} and in ecological resilience restoration \cite{ma2021universality}. Here, stochasticity acts as a ``safety valve'' preventing the build--up of latent hysteretic inertia.

\paragraph{3. Sentinel monitoring and predictive homeostasis.}  
Building on the above, we formalised an internal ``sentinel'' module implementing a simple early-warning strategy grounded in critical slowing down \cite{carpenter2006rising, scheffer2001catastrophic}. In our implementation, the monitored metric is the \emph{average power per cell} $\overline{P}_{\mathcal{S}}(t)$ (Eq.~\ref{eq:avg_pow_cell}) computed over a chosen subset of sites $\mathcal{S}$ and a sliding time window. As $\alpha$ approaches the phase boundary, $\overline{P}_{\mathcal{S}}(t)$ often exhibits a gradual \emph{increasing trend}, reflecting the amplified fluctuations characteristic of systems near loss of stability.  

To quantify this trend, the sentinel computes Kendall’s $\tau_K$ rank correlation coefficient between $\overline{P}_{\mathcal{S}}(t)$ and time $t$ over the most recent $W$ points. Large positive $\tau_K$ values indicate a significant upward trend, consistent with an impending transition, while values near zero suggest no monotonic trend. By thresholding $\tau_K$, the module discriminates between slow approaches to instability (``approaching parturition'') and trajectories that remain safely away from the boundary (``quiescent'') (Fig.~\ref{fig:avgpower-eval}). When $\tau_K$ exceeds the decision threshold, the controller halts the ongoing decrease of $\alpha$ or applies a corrective hormonal pulse.  

Notably, uncontrolled cells convey more predictive information than controlled ones: because their dynamics are less suppressed, they respond more visibly to early signs of instability, yielding higher classification accuracy. This architecture parallels the broader concept of predictive (anticipatory) homeostasis, whereby physiological controllers exploit naturally occurring fluctuations to forecast and pre-empt critical state changes \cite{simor2023sleep, deans2021biological}.  
While direct identification of controlled versus uncontrolled myocytes is currently impossible in vivo, this prediction can be confronted indirectly with empirical findings from the electrohysterography (EHG) literature. Several studies comparing electrode configurations report that signals recorded near the uterine fundus—particularly in the upper left or right regions—carry disproportionate information for distinguishing uterine contractions from background activity or for classifying pregnancy versus labor. For example, classification accuracy has been shown to depend strongly on electrode location, with fundal electrodes consistently outperforming lower or cervical ones across different datasets and methods~\cite{hao2019application, hao2021effect, moslem2011classification}. Related work on multichannel fusion further demonstrates that not all spatial regions contribute equally to labor discrimination, and that a small subset of electrodes can dominate predictive performance~\cite{rabotti2015propagation, young2007myocytes, moslem2012classification}.

From the perspective of our model, these findings are consistent with the interpretation that fundal regions behave as weakly controlled sites, where synchronized activity and increased variance emerge earlier and more robustly. This interpretation gains biological plausibility from independent evidence that progesterone-mediated suppression of contractility is spatially heterogeneous within the uterus. While circulating progesterone levels remain high throughout pregnancy, multiple studies indicate region-specific modulation of progesterone receptor (PR) signaling, particularly in the myometrium. In human fundal myometrium, labor and late gestation are associated with reduced expression of PR co-activators (such as SRC-2, SRC-3, and CBP), increased expression of inhibitory PR isoforms (e.g., PR-C), and up-regulation of PR-associated corepressors, all of which diminish progesterone responsiveness without requiring a change in hormone concentration~\cite{condon2006up}. These changes are less pronounced or absent in the lower uterine segment, which plays a distinct mechanical and functional role during parturition.

\paragraph{Implications for preterm birth.}
Preterm birth is widely recognized as a heterogeneous condition involving multiple tissues and biological pathways, with no single dominant mechanism accounting for most cases \cite{vidal2022spontaneous}. Although a full aetiological theory is beyond our scope, the model provides a unifying dynamical language for several hypotheses concerning preterm parturition. In this phase--space view, preterm birth corresponds to a premature loss of control that precipitates sustained global oscillations. Two generic modes of control failure can be distinguished. The first, illustrated explicitly in our simulations, arises from misidentification of the stability boundary $\alpha_c(n_c,D)$, particularly when the controllable fraction $n_c$ is low, for example due to structural or endocrine constraints such as regional progesterone resistance \cite{condon2006up,ji2024exploring}. A second, conceptually distinct failure mode occurs when the control system correctly identifies its working boundaries but is unable to re-engage effectively following transient excursions. In such scenarios, control is lost not because its target is misestimated, but because recovery or reactivation is impaired, allowing brief contractions to nucleate sustained global activity. Other biologically plausible pathways---such as rapid increases in intercellular coupling (increasing $D$) or changes in noise structure---remain speculative at this stage and constitute targets for new analyses. 

\paragraph{Limitations and future directions.}
Several simplifying assumptions delimit our conclusions. Geometry was restricted to a homogeneous square lattice without mechanical feedback: in vivo uterine tissue exhibits anisotropy, curvature, and stretch–activated conductances \cite{fang2021anisotropic,fidalgo2022effect}. The controller was linear and memoryless aside from a tunable relaxation time; endocrine networks include multiple interacting non-linear loops and finite propagation delays. We lumped heterogeneous hormonal influences into a single scalar $\alpha$ and modeled spatial heterogeneity via a bimodal distribution. The cost functional was treated implicitly: an explicit optimization problem combining a penalty for oscillatory power with metabolic expenditure \cite{bechhoefer2021control} could yield quantitative predictions for $\alpha(n_c)$ trajectories and clarify trade–offs under pathological perturbations. Moreover, the early-warning analysis implemented \emph{detection only}. A natural extension is to introduce an actuation channel (“hormonal pulse”) that modulates the $\alpha$-dynamics once a trigger is declared, for example by adding a short input $u(t)$,
\[
\dot\alpha \;=\; -r \;+\; \kappa\,u(t)\,,
\]
with $u(t)$ a brief pulse initiated when the EWS crosses a threshold (equivalently, applying a transient offset $\alpha(t)\!\to\!\alpha(t)+\Delta\alpha\,e^{-(t-t^\ast)/\tau_p}$). This closed-loop variant would allow us to test whether timely pulses can arrest or reverse the drift toward oscillations while quantifying the associated control costs. Finally, the early warning analysis considered a single aggregated metric, which might be overly simple even for biological realism.

\section{Conclusions}
\label{sec:conc}
We have presented a dynamical model that reframes the pregnant uterus as a sparsely controlled network of coupled oscillators, maintained near a stability boundary through cost-aware feedback. This perspective addresses three longstanding puzzles in the literature. First, it provides a unifying dynamical language for preterm birth, not by positing a single aetiology, but by interpreting heterogeneous biological insults as distinct failure modes of a common control architecture, all capable of precipitating a premature loss of quiescence. Second, it offers a mechanistic reinterpretation of pre-labour contractions—such as Alvarez and Braxton--Hicks waves—not as preparatory ``training'' events, but as natural noise-driven excursions in a system operating close to instability, whose occurrence reflects control efficiency rather than muscular conditioning. Third, the model resolves the robustness--flexibility paradox by showing how long-term suppression of global synchronization can coexist with rapid, decisive activation at term through the release of a latent fast dynamical mode once control is withdrawn.

Beyond these conceptual contributions, the framework rationalizes several empirical observations and suggests concrete directions for analysis. In particular, it clarifies why uterine electrical activity should be spatially heterogeneous, with weakly controlled regions acting as sentinels where fluctuations and early-warning signals emerge first, consistent with reports that specific electrode locations carry disproportionate diagnostic information. It also predicts that uterine activity should organize along a continuum of event magnitudes as control weakens, although current EHG recordings lack such spatial resolution.

More broadly, the framework shifts attention from isolated molecular or tissue-specific mechanisms toward the dynamical organization of control itself. By articulating how robustness, flexibility, noise, and heterogeneity interact to govern the timing of labour, this work offers a coherent foundation for integrating diverse empirical findings and for guiding future analyses of uterine dynamics, without presupposing a unique biological trigger or clinical intervention.

\vskip 0.5cm
\noindent{\bf Acknowledgements}: This work is partially supported by the National Natural Science Foundation of China (Grant no. T2350710802), Shenzhen Science and Technology Innovation Commission Project (grant No. K23405006), and the Center for Computational Science and Engineering at Southern University of Science and Technology.
\\
A.S. acknowledges support by the Dutch Econophysics Foundation (Stichting Econophysics Leiden, the Netherlands) and by the European Union - NextGenerationEU - National Recovery and Resilience Plan (Piano Nazionale di Ripresa e Resilienza, PNRR), projects ``SoBigData.it - Strengthening the Italian RI for Social Mining and Big Data Analytics'', Grant IR0000013 (n. 3264, 28/12/2021) (\url{https://pnrr.sobigdata.it/}), and ``Reconstruction, Resilience and Recovery of Socio-Economic Networks'' RECON-NET EP\_FAIR\_005 - PE0000013 ``FAIR'' - PNRR M4C2 Investment 1.3. He also acknowledges support by the European Union under the scheme HORIZON-INFRA-2021-DEV-02-01 – Preparatory phase of new ESFRI research infrastructure projects, Grant Agreement n.101079043, “SoBigData RI PPP: SoBigData RI Preparatory Phase Project”.
\\
G.M.F. conducted most of this research while affiliated with the Department of Ecology and Evolutionary Biology at Princeton University. G.M.F. acknowledges support from the Swiss National Science Foundation (grant P500PS-211064), a gift from William H. Miller III, and the Schmidt DataX Fund at Princeton University, made possible through a major gift from Schmidt Futures.

\bibliography{sample}
\bibliographystyle{unsrt}

\clearpage
\section*{Supplementary Material}
\addcontentsline{toc}{section}{Supplementary Material}

\setcounter{section}{0}
\renewcommand{\thesection}{A\arabic{section}}
\renewcommand{\thesubsection}{A\arabic{section}.\arabic{subsection}}
\renewcommand{\thesubsubsection}{A\arabic{section}.\arabic{subsection}.\arabic{subsubsection}}
\beginsupplement
\section{Additional analyses for the Single cell case}
\subsection{Bifurcation diagram for a single cell}
Let us rewrite here for convenience the system of equations for an isolated cell in absence of noise.
\begin{equation}
\label{eqApp:sys1dPRIMA}
\left\{
\begin{aligned}
\frac{dV}{dt} \; &= \; A\,V\,\bigl(V - \omega\bigr)\,\bigl(1 - V\bigr)
\;-\; g
\;+\; n_{p}\,C_r\,\bigl(V_{p} - V\bigr)
\;-\; \mu
\\[6pt]
\frac{dg}{dt} \; &= \; \epsilon \,\bigl(V - g\bigr)
\\[6pt]
\frac{dV_{p}}{dt} \; &= \; K\,\bigl(V_p^{R} - V_{p}\bigr)-C_r\bigl(V_{p} - V\bigr)
\\[6pt]
\frac{d\mu}{dt} \; &=  \gamma\,\bigl(\alpha\,V\;-\;\mu \bigr)
\end{aligned}
\right.
\end{equation}
where the target potential has been set to $V_T=0$. The parameters are $A=3$, $\omega=0.2$, $\epsilon=0.08$, $K=0.25$, $V_p^{R}=1.5$, $\gamma=1$, set so that the system has a unique steady state $(V^*, g^*, V_p^*, \mu^*)$. Define
\begin{equation}
\label{eqSI:fv}
f_V(V) \;=\; \frac{\partial}{\partial V}\Big[A\,V\,(V-\omega)\,(1-V)\Big]
\;=\; A\bigl(-3V^2 + 2(1+\omega)V - \omega\bigr).
\end{equation}
The Jacobian $J$ at the steady state $(V^*,g^*,V_p^*,\mu^*)$ is
\begin{equation}
\label{eqApp:J}
J \;=\;
\begin{pmatrix}
f_V(V^*) - n_p C_r & -1 & n_p C_r & -1\\[2pt]
\epsilon & -\epsilon & 0 & 0\\[2pt]
C_r & 0 & -(K+C_r) & 0\\[2pt]
\gamma \alpha & 0 & 0 & -\gamma
\end{pmatrix}.
\end{equation}

\begin{figure}[htbp!]
  \centering
  \includegraphics[width=0.92\linewidth]{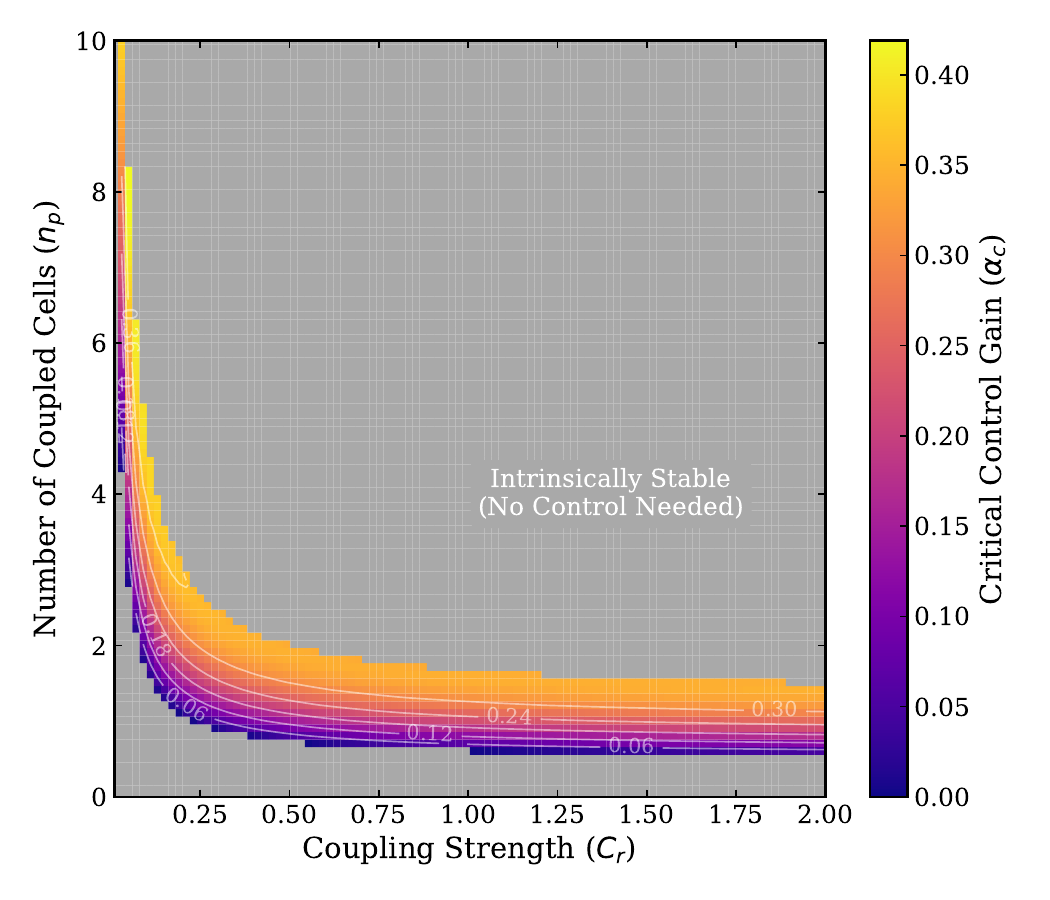}

  \caption{%
    Single--cell bifurcation diagram.
  Heat map of the critical gain $\alpha_c(C_r,n_p)$ for the isolated system \eqref{eqApp:sys1dPRIMA}.
  Colours show the value of $\alpha_c$ obtained by solving $\Lambda_{\max}(\alpha;C_r,n_p)=0$ (largest real part of the Jacobian eigenvalues equals zero); thin white contours aid visual reading.
  Grey indicates parameter combinations that are already linearly stable at $\alpha=0$ (no control needed).
  Parameters: $A=3$, $\omega=0.2$, $\epsilon=0.08$, $K=0.25$, $V_p^R=1.5$, $\gamma=1$.
  }
  \label{fig:pd_single_cell}
\end{figure}

\paragraph{Hopf bifurcation test.}
Let $\chi(\lambda)=\det(\lambda I - J)=\lambda^4+a_1\lambda^3+a_2\lambda^2+a_3\lambda+a_4$ be the characteristic polynomial. The equilibrium is linearly stable if the Routh--Hurwitz inequalities hold:
\[
a_1>0,\quad a_2>0,\quad a_3>0,\quad a_4>0,
\qquad
\Delta_3 \;\equiv\; a_1 a_2 a_3 \;-\; a_3^2 \;-\; a_1^2 a_4 \;>\; 0.
\]
A generic Hopf crossing occurs when $\Delta_3=0$ while the other inequalities remain satisfied. In our computations, we use an equivalent and robust condition based on the spectral abscissa,
\[
\Lambda_{\max}(\alpha;C_r,n_p)
\;\equiv\;
\max_i \Re\,\lambda_i\!\left(J(\alpha;C_r,n_p)\right),
\]
and define $\alpha_c$ as the solution of $\Lambda_{\max}(\alpha_c;C_r,n_p)=0$. Fig. \ref{fig:pd_single_cell} shows the dependence of $\alpha_c$ on the number of passive cells $n_p$ and the coupling between active and passive cells $C_r$.
\subsection{Control in the presence of observation delay}
\label{sec:delay_control}

In Sec.~\ref{secsub:scellcontrol} of the main text, we analyzed the dynamics of a single controlled excitable cell in the absence of both noise ($\chi_v = 0$) and observation delay ($\tau = 0$). In that setting, the control term $\mu(t)$ is driven instantaneously by the current state of the system.  

We now extend the analysis to the case $\tau > 0$, where the control responds to the \emph{past} state of the system. Specifically, in the last equation of system \eqref{eq:sys1dPRIMA}
\begin{equation}
\frac{d\mu}{dt} = \gamma\bigl( \alpha\, r(t - \tau) - \mu(t) \bigr),
\label{eq:mu_delay}
\end{equation}
the recovery variable $r$ (or equivalently the measured voltage) is evaluated at time $t-\tau$. The resulting model is a system of \emph{delay differential equations} (DDEs).  

Intuitively, finite delays in the feedback loop act like a phase lag: the control acts on outdated information, and for sufficiently large $\tau$ this lag can amplify oscillations rather than suppress them. The goal of this section is to determine, for each $\alpha$ (with $\gamma=1$ as in the main text), the \emph{maximum} delay $\tau_{\mathrm{max}}$ the system can tolerate before the steady state loses stability.

\paragraph{Linearization.}
Let $(V^*, g^*, V_p^*, \mu^*)$ denote the fixed point of Eqs.~\eqref{eq:sys1dPRIMA}--\eqref{eq:mu_delay}, and let $\delta V(t) = V(t) - V^*$, $\delta g, \delta V_p, \delta\mu$, denote small perturbations. The linearized dynamics can be written in the compact DDE form
\begin{equation}
\dot{\mathbf{x}}(t) = A_0 \,\mathbf{x}(t) + A_1\, \mathbf{x}(t - \tau),
\label{eq:lin_dde}
\end{equation}
where $\mathbf{x} = (\delta V, \delta g, \delta V_p, \delta\mu)^\mathsf{T}$.  

The Jacobian $A_0$ collects the terms without delay, while $A_1$ contains only the delayed coupling introduced via $\delta V(t-\tau)$ in Eq.~\eqref{eq:mu_delay}. Explicitly, for general $n_p$ and $C_r$ they read
\[
A_0 \;=\;
\begin{pmatrix}
f_V(V^*) - n_p C_r & -1 & \;\; n_p C_r \;\; & -1\\[4pt]
\epsilon & -\epsilon & 0 & 0\\[4pt]
C_r & 0 & -(K+C_r) & 0\\[4pt]
0 & 0 & 0 & -\gamma
\end{pmatrix},
\qquad
A_1 \;=\;
\begin{pmatrix}
0 & 0 & 0 & 0\\
0 & 0 & 0 & 0\\
0 & 0 & 0 & 0\\
\gamma\alpha & 0 & 0 & 0
\end{pmatrix},
\]
where $f_V(V)$ is defined in Eq. \ref{eqSI:fv}.

\paragraph{Characteristic equation.}
For a DDE of the form \eqref{eq:lin_dde}, one seeks exponential solutions $\mathbf{x}(t) = \mathbf{u} e^{\lambda t}$, $\mathbf{u}\neq 0$. Substitution yields the \emph{characteristic equation}
\begin{equation}
\det\big[ \lambda I - A_0 - A_1 e^{-\lambda\tau} \big] = 0.
\label{eq:char_eq}
\end{equation}
This is a transcendental equation in $\lambda$ due to the term $e^{-\lambda\tau}$, implying an infinite spectrum of eigenvalues.  

\paragraph{Hopf condition.}
The boundary of stability is reached when a pair of complex-conjugate eigenvalues crosses the imaginary axis: $\lambda = i\theta$, $\theta > 0$. Substituting into \eqref{eq:char_eq} and separating real and imaginary parts yields a system of two equations for $\theta$ and $\tau$:
\begin{equation}
\Re\Big\{ \det\big[ i\theta I - A_0 - A_1 e^{-i\theta\tau} \big] \Big\} = 0, 
\quad
\Im\Big\{ \det\big[ i\theta I - A_0 - A_1 e^{-i\theta\tau} \big] \Big\} = 0.
\label{eq:hopf_conditions}
\end{equation}
For each $\alpha$, these equations can be solved numerically to yield $\theta$ and the corresponding critical delay $\tau_{\mathrm{max}}$. This provides the dependence of the maximum delay time $\tau_{\mathrm{max}}(\alpha)$ for a given control strength $\alpha$.

\begin{figure}[htbp!]
  \centering
  \includegraphics[width=0.92\linewidth]{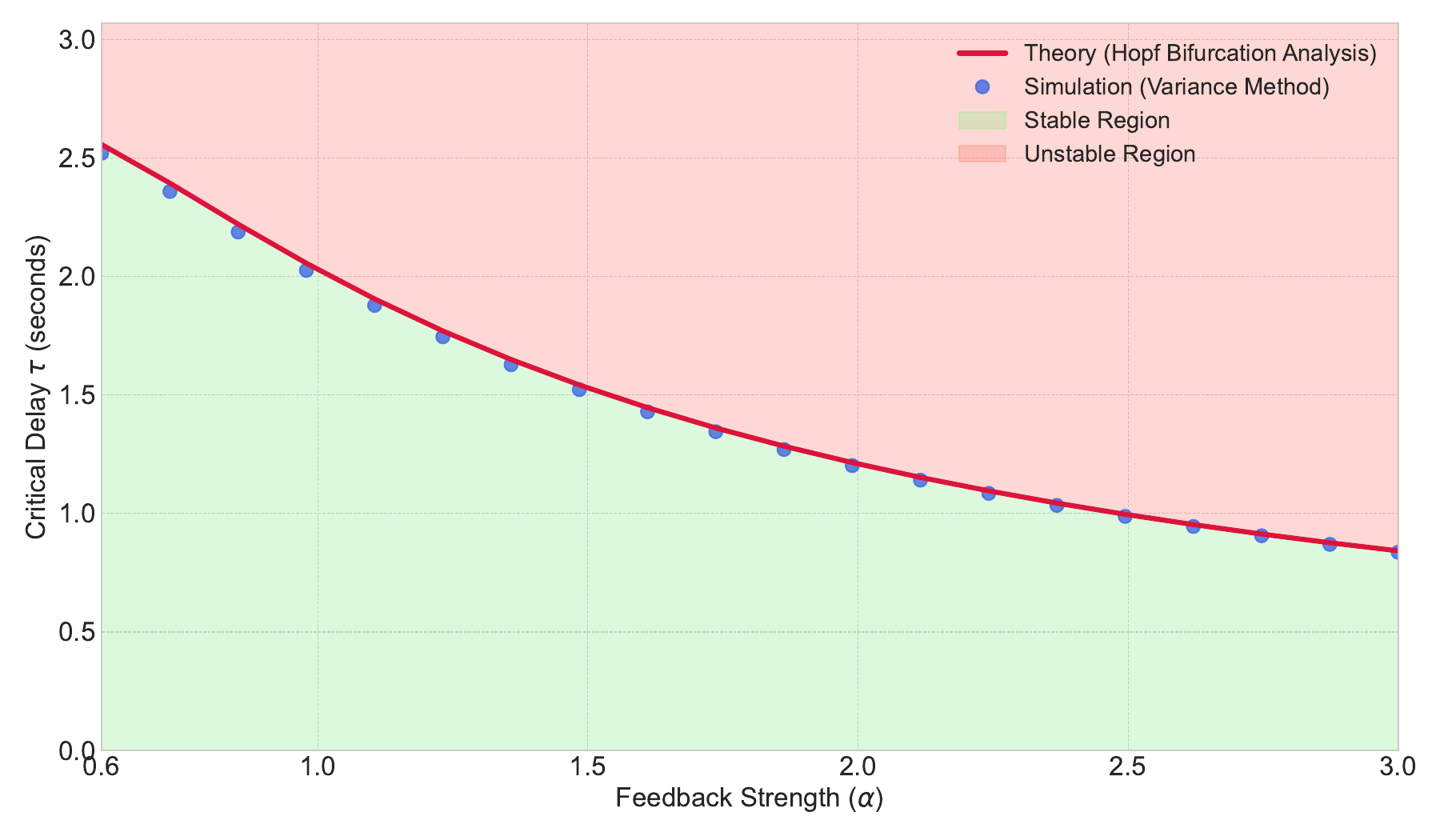}

  \caption{%
    Stability boundary in the feedback-delay parameter space. The maximum delay $\tau$ for which stability remains is plotted as a function of the feedback strength $\alpha$. The solid red line represents the theoretically derived Hopf bifurcation boundary , determined via linear stability analysis of the system's fixed point leading to equations (\ref{eq:hopf_conditions}). The blue circles are obtained from direct numerical simulations of the full nonlinear DDE system, where the onset of instability was detected by measuring the variance of the voltage $V(t)$. The shaded green area below the boundary indicates the region of stable steady-state behavior, while the shaded salmon area above it marks the region where the system exhibits sustained oscillations. The excellent agreement between the theoretical prediction and the numerical results validates the linear stability analysis for determining the onset of delay-induced instability in the model.
  }
  \label{fig:tau_delay}
\end{figure}

\paragraph{Numerical verification.}
We have verified the function $\tau_{\mathrm{max}}(\alpha)$ obtained from solving (\ref{eq:hopf_conditions}) by comparing with direct numerical simulation of the DDE using the \texttt{jitcdde} Python library. For each $\alpha$, we increased $\tau$ in small increments and monitored the long-term dynamics of $V(t)$. The maximal delay before the onset of sustained oscillations was recorded as $\tau_{\mathrm{max}}$. The analytical and numerical results showed excellent agreement (Fig.~\ref{fig:tau_delay}), confirming the validity of the linear stability analysis.

Figure~\ref{fig:tau_delay} shows a clear trade-off: stronger feedback (larger $\alpha$) is markedly more sensitive to delay. Although not highlighted in the main text, this offers an additional rationale for choosing a modest control gain: a smaller $\alpha$ not only lowers energetic cost (Fig.~\ref{figDS:newfig}) but also enlarges the delay margin, making the controller more robust to observation lags.

\clearpage
\section{Additional analyses of the spatially extended system: static bifurcation in absence of noise}
\subsection{Variability of the critical control gain  $\alpha_c$}
\label{sec:si_onset_variability}

Figure~\ref{figDS:main} in the main text shows that the critical gain $\alpha_c$ required to suppress oscillations exhibits substantial variability from realisation to realisation when only a small fraction $n_c$ of cells is controlled. Here we quantify and explain this variability. As in the main text, for each realization (fixed $D$, $n_c$, seed), we define $\alpha_c$ as the smallest $\alpha$ (found by bisection) that suppresses oscillations on the lattice, so that the fraction of oscillatory sites 
drops to zero at long times. We collect the different values of $\alpha_c$ obtained over different numerical seeds controlling different realisations of the system at fixed $(D,n_c)$ and obtain the coefficient of variation $\mathrm{CV}=\sigma(\alpha_c)/\langle \alpha_c\rangle$ defined as the ratio of the standard deviation to the average of the statistical ensemble of obtained $\alpha_c$'s. 

\begin{figure}[htbp]
  \centering
  \includegraphics[width=0.92\linewidth]{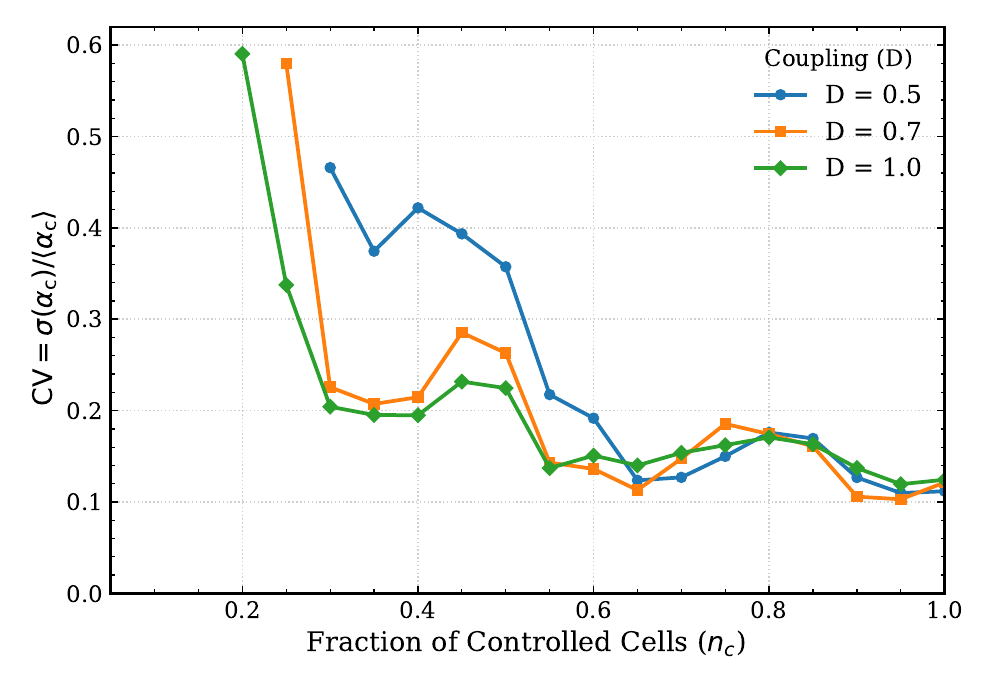}
  \caption{
  Coefficient of variation $\mathrm{CV}(\alpha_c)$ over 10 statistically equivalent realisations of the system as a function of $n_c$ for the three diffusivities $D\in\{0.5,0.7,1.0\}$. In other words, each point is obtained by calculating the standard deviation and the average over 10 seeds at fixed $(D,n_c)$. Variability is largest at low $n_c$, consistent with an extreme-value mechanism in which a single weakly coupled uncontrolled island sets the onset.}
  \label{fig:cv_alpha_onset}
\end{figure}

Figure~\ref{fig:cv_alpha_onset} shows that $\mathrm{CV}(\alpha_c)$ is largest at low $n_c$ and decreases as control coverage increases. This trend is observed at all three diffusivities $D$, with overall variability depending moderately on $D$.

\paragraph{Largest cluster dominance at low $n_c$; boundary quenching at high $n_c$}
At small fractions $n_c$ of excitable cells that are controlled (low coverage), the onset is governed by a largest-cluster effect, which can be formulated within extreme value theory: global oscillations persist as long as at least one sufficiently large uncontrolled cluster of connected cells remains self-sustained despite the leakage imposed by diffusive control. We therefore focus on the set
\[
U \;=\;\{\,i:\ i\notin\mathcal{C},\ n_{p,i}=1\,\}
\]
of sites that are \emph{uncontrolled} and carry a single passive load\footnote{Recall that we sample $n_p$ for each cell from a poisson distribution with mean $0.7$, so about 35\% of cells have $n_p=1$.}. For our parameters ($C_r=1$), an isolated $n_p{=}1$ unit is intrinsically oscillatory (see Fig. \ref{fig:pd_single_cell}, making these sites the natural seeds for persistence of oscillatory behavior at low $n_c$.

\begin{figure}[ht]
  \centering
  \begin{subfigure}[b]{0.48\textwidth}
    \centering
    \includegraphics[width=\linewidth]{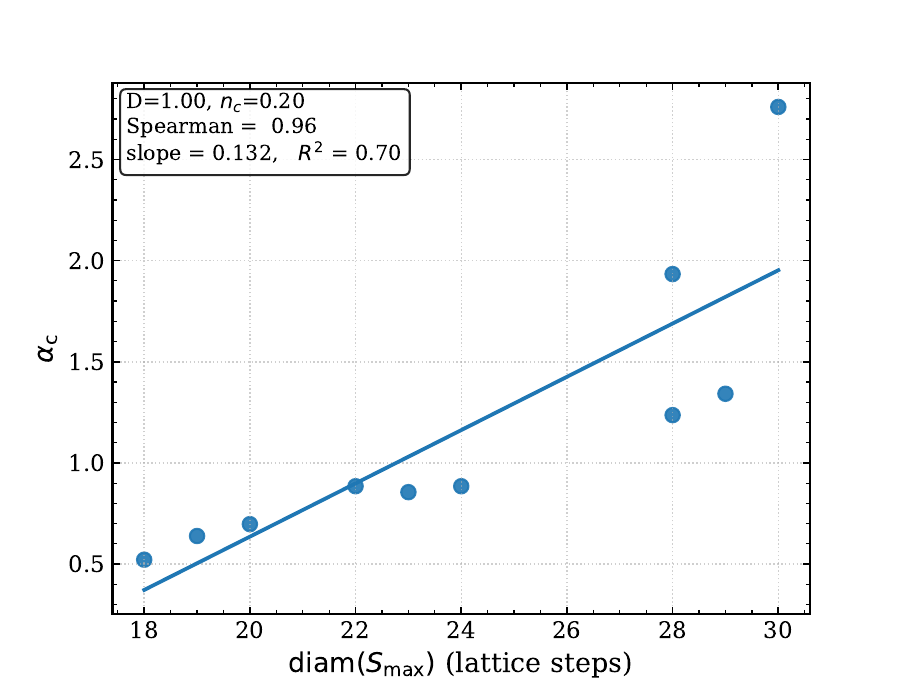}
    \caption{$D=1.0$, $n_c=0.20$ (low coverage).}
    \label{fig:onset_scatter_low}
  \end{subfigure}\hfill
  \begin{subfigure}[b]{0.48\textwidth}
    \centering
    \includegraphics[width=\linewidth]{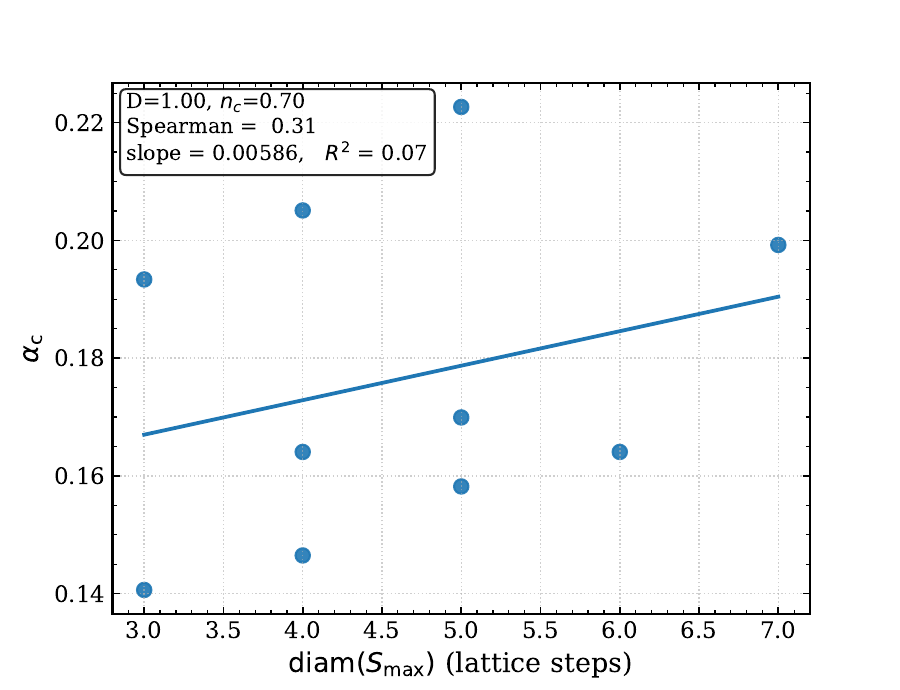}
    \caption{$D=1.0$, $n_c=0.70$ (high coverage).}
    \label{fig:onset_scatter_high}
  \end{subfigure}
  \caption{
Dependence of the critical control gain $\alpha_c$ as a functin of the geodesic diameter $\mathrm{diam}(S_{\max})$ of the largest uncontrolled oscillator cluster, for two representative control coverages $n_c$ for $D=1.0$.
\emph{Left} ($n_c=0.2$): clear positive trend indicating that larger islands are harder to extinguish and require a larger critical control gain $\alpha_c$.
\emph{Right} ($n_c=0.7$): the trend disappears, consistent with the absence of a single dominant large cluster driving the persistence of oscillations at onset.
Each panel shows the OLS best-fit line, with slope, $R^2$, and Spearman’s~$\rho$ reported in the inset.}
  \label{fig:ab}
\end{figure}

Let $S_{\max}\subset U$ denote the largest connected component (using the criterion of 4-neighbour adjacency on the $L\times L$ torus), and write $\mathrm{diam}(S_{\max})$ for its graph-theoretical diameter,
\[
\mathrm{diam}(S_{\max})=\max_{x,y\in S_{\max}}\mathrm{dist}_{S_{\max}}(x,y).
\]
In the panel corresponding to low coverage (small $n_c$, Fig.~\ref{fig:onset_scatter_low}), one can observe a clear \emph{positive} correlation between the critical control gain $\alpha_c$ and $\mathrm{diam}(S_{\max})$: larger islands (which also tend to have larger areas) require a larger control amplitude to quench. This is consistent with a size-dominated balance, where the internal drive scales with area while diffusive leakage scales with boundary length, making larger islands harder to extinguish.

By contrast, in the high coverage panel (high $n_c$, Fig.~\ref{fig:onset_scatter_high}) the correlation essentially vanishes. Here $S_{\max}$ shrinks and is strongly exposed to controlled neighbours; no single island dominates the onset and geometric predictors lose explanatory power\footnote{Other connectivity descriptors—e.g.\ the algebraic connectivity $\lambda_2$ of the Neumann Laplacian on the induced subgraph of $S_{\max}$—show analogous trends at low $n_c$ and similarly fade at high $n_c$; see, e.g., classical discussions of $\lambda_2$ as a bottleneck measure \cite{fiedler1973algebraic}.}.

\subsection{Robust control onset estimator ($\alpha_{50}$)}
\label{sec:bulk}
\paragraph{Definition and estimation.}
Because the critical control gain at the onset of oscillations varies substantially across realizations (i.e. across different seeds, at fixed $(D,n_c)$), we define a consistent control onset proxy, denoted $\alpha_{50}$. For each realization, we compute $f_{\mathrm{osc}}(\alpha)$ (Eq.\ref{eq:metrics_control}) by performing simulation with parameters scanning a grid of $\alpha\in[0,\alpha_{\mathrm{c}}]$, performing independent runs for each $\alpha$. We then fit the two-parameter logistic
\[
f_{\mathrm{osc}}(\alpha)=\frac{1}{1+\exp\!\big(s(\alpha-\alpha_{50})\big)},\qquad s>0.
\]
and define $\alpha_{50}$ as the inflection point of the fitted curve, which is defined for each realisation. This provides a reproducible definition of onset across heterogeneous seeds.

\begin{figure}[htbp]
  \centering
  \includegraphics[width=0.92\linewidth]{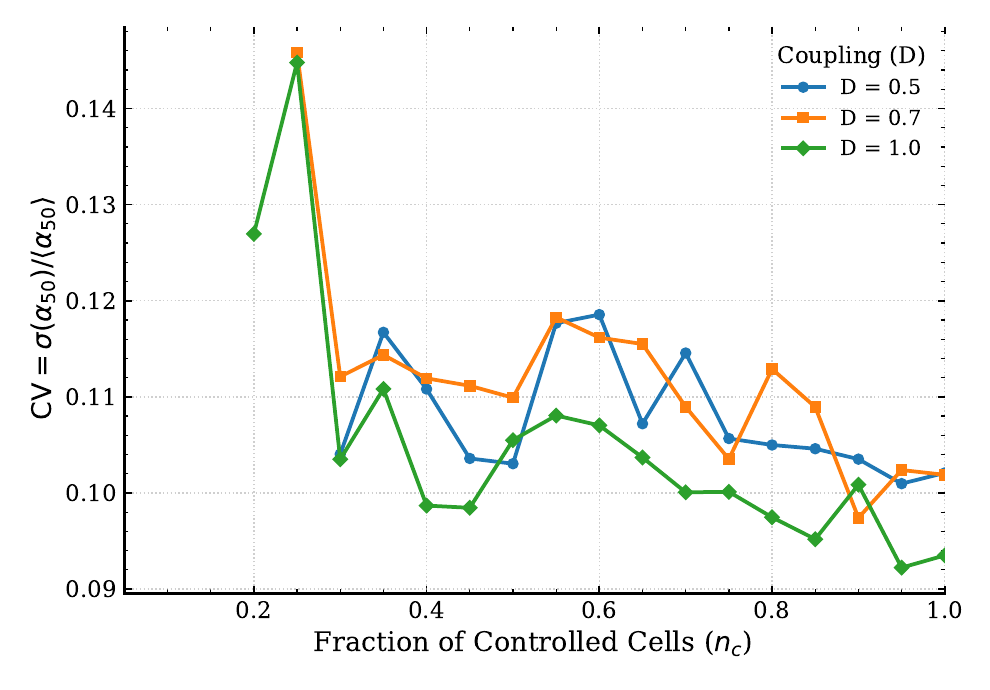}
  \caption{Variability of inflection control threshold as a function of control coverage $n_c$.
  Coefficient of variation $\mathrm{CV}(\alpha_{50})$ as a function of $n_c$ for the three diffusivities $D\in\{0.5,0.7,1.0\}$. Each point corresponds to an average over 10 realisations at fixed $(D,n_c)$. Compare with variability of $\alpha_c$ in Fig. \ref{fig:cv_alpha_onset}.}
  \label{fig:cv_alpha_50}
\end{figure}

\paragraph{Variability and interpretation.}
We summarise the per-realisation estimates $\alpha_{50}$ by their mean and standard deviation, and report the coefficient of variation $\mathrm{CV}=\sigma(\alpha_{50})/\langle \alpha_{50}\rangle$ as a function of $n_c$ and $D$ (Fig.~\ref{fig:cv_alpha_50}). Empirically, $\mathrm{CV}(\alpha_{50})$ is consistently small for all values of $n_c$ for all three $D$ values (compare Fig.~\ref{fig:cv_alpha_50} with Fig.~\ref{fig:cv_alpha_onset}). Unlike the critical control gain $\alpha_{\mathrm{c}}$ (which, at low $n_c$, is controlled by the “worst” uncontrolled island), the inflection $\alpha_{50}$ is a system-wide property: it marks the gain at which roughly half the lattice ceases oscillating. Consequently, it is much less sensitive to any single cluster and exhibits markedly smaller realisation-to-realisation spread, which is consistent with the property of self-averaging when many subdomains contribute to $f_{\mathrm{osc}}$.

\paragraph{Agreement with heterogeneous mean-field theory.}
We compare the empirical $\alpha_{50}$ with the Hopf threshold $\alpha_{\mathrm{c}}^{(2\mathrm{cell})}(D,n_c)$ of the two-cell heterogeneous mean-field model (Sec.~\ref{sec:meanfield}). Briefly, the model couples a “controlled’’ and a “free’’ unit with effective weights $D(1-n_c)$ and $Dn_c$. We locate the Hopf point by solving $\max\Re\lambda(\alpha,D,n_c)=0$ at the fixed point. For each $D$, we compare the empirical mean $\langle\alpha_{50}\rangle$ (shaded band $=\pm$ one s.d.) and the theoretical value $\alpha_{\mathrm{c}}^{(2\mathrm{cell})}(D,n_c)$ obtained as function of $n_c$. For all values of $D$, the system-wide inflection control point measured in the spatial system closely tracks the mean-field prediction, with small variability over different realisations.

As a check, we also form a mean estimate $\alpha_{50}^{\mathrm{pool}}$ by first averaging $f_{\mathrm{osc}}(\alpha)$ over different realisations for each $\alpha$ and then fitting the same two-parameter logistic to the average curve (Fig. \ref{fig:foscvsalpha}). As expected for a system-wide transition, $\alpha_{50}^{\mathrm{pool}}$ is virtually identical to $\alpha_{\mathrm{c}}^{(2\mathrm{cell})}$, reflecting reduced statistical fluctuations when averaging the full $f_{\mathrm{osc}}(\alpha)$ profile before fitting. In this system-wide controlled regime, the property of self-averaging holds.

\begin{figure}[htbp]
  \centering
  \includegraphics[width=0.92\linewidth]{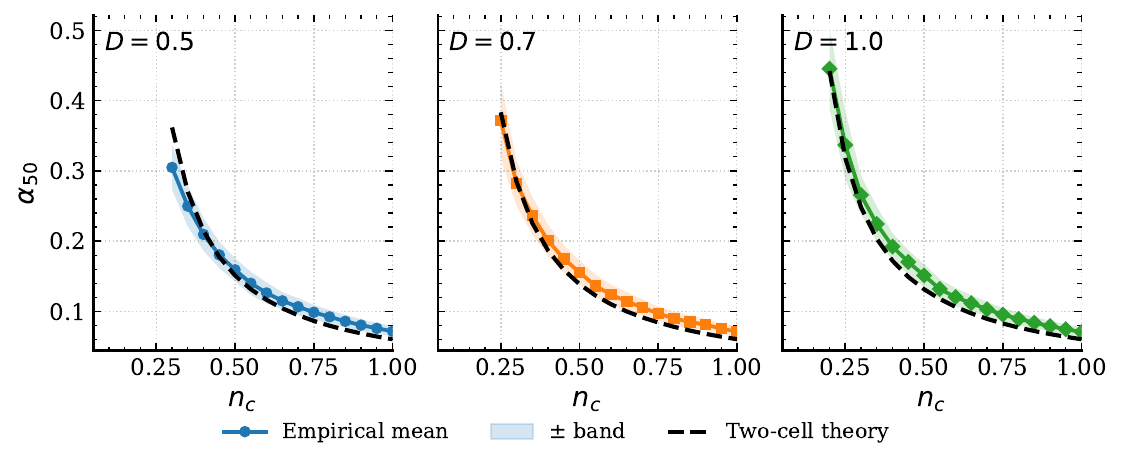}
  \caption{System-wide inflection control value $\alpha_{50}$ as a function of control coverage $n_c$: simulations vs.\ theory.
Each panel corresponds to a fixed coupling $D\in\{0.5,\,0.7,\,1.0\}$ and shows, for every $n_c$, the empirical mean $\langle\alpha_{50}\rangle$ obtained for 10 seeds (markers) with a shaded band denoting $\pm1$ s.d., together with the two-cell heterogeneous mean-field prediction $\alpha_c^{(2\mathrm{cell})}(D,n_c)$ (dashed line). 
The value of $\alpha_{50}$ for each realisation is obtained by fitting a two-parameter logistic to $f_{\mathrm{osc}}(\alpha)$; the theorical curve is the Hopf threshold of the two-unit model with effective couplings $D(1-n_c)$ and $Dn_c$ (computed by solving $\max\Re\lambda(\alpha,D,n_c)=0$ at the fixed point). }

  \label{fig:cv_alpha_smallmultip}
\end{figure}

\subsection{Two-cell heterogeneous mean-field model}
\label{sec:meanfield}
To gain analytical insight into the system’s dynamics, we reduce the full spatial model to a coarse-grained, two-element heterogeneous mean-field approximation. This simplified model consists of two FitzHugh–Nagumo units representing the membrane potentials of a “controlled” and a “free” (uncontrolled) cell, each coupled to a passive element and to each other through effective diffusive coupling. The dynamics are governed by the following system of equations:
\begin{align}
\begin{split}
\label{eq:1cellnocontrol}
 &\frac{dV_{e,1}}{dt}=AV_{e,1}(V_{e,1}-\eta)(1-V_{e,1})-g_1+n_pC_r(V_{p,1}-V_{e,1})-\mu_1 +D(1-n_c)(V_{e,2}-V_{e,1})\\
 &\frac{dg_1}{dt}=\epsilon(V_{e,1}-g_1)\\
 &\frac{dV_{p,1}}{dt}=K(V_p^R-V_{p,1})-C_r(V_{p,1}-V_{e,1})\\
 &\frac{d\mu_1}{dt}=\gamma(\alpha V_{e,1}-\mu_1)\\
 &\frac{dV_{e,2}}{dt}=AV_{e,2}(V_{e,2}-\eta)(1-V_{e,2})-g_2+n_pC_r(V_{p,2}-V_{e,2})+Dn_c(V_{e,1}-V_{e,2})\\
 &\frac{dg_2}{dt}=\epsilon(V_{e,2}-g_2)\\
 &\frac{dV_{p,2}}{dt}=K(V_p^R-V_{p,2})-C_r(V_{p,2}-V_{e,2})\\
\end{split}
\end{align}
As mentioned in Section \ref{sec:bulk}, the value $\alpha_c^{(2\mathrm{cell})}$ at which a hopf bifurcation occurs is in excellent agreement with the inflection point of the fraction of oscillating sites $f_o=\frac{1}{L^2}\sum_{i,j}H(p_{ij}-p_{th})$ in the spatially extended system.
\begin{figure}[H]
    \centering
    \subfloat[]{
        \includegraphics[width=\textwidth]{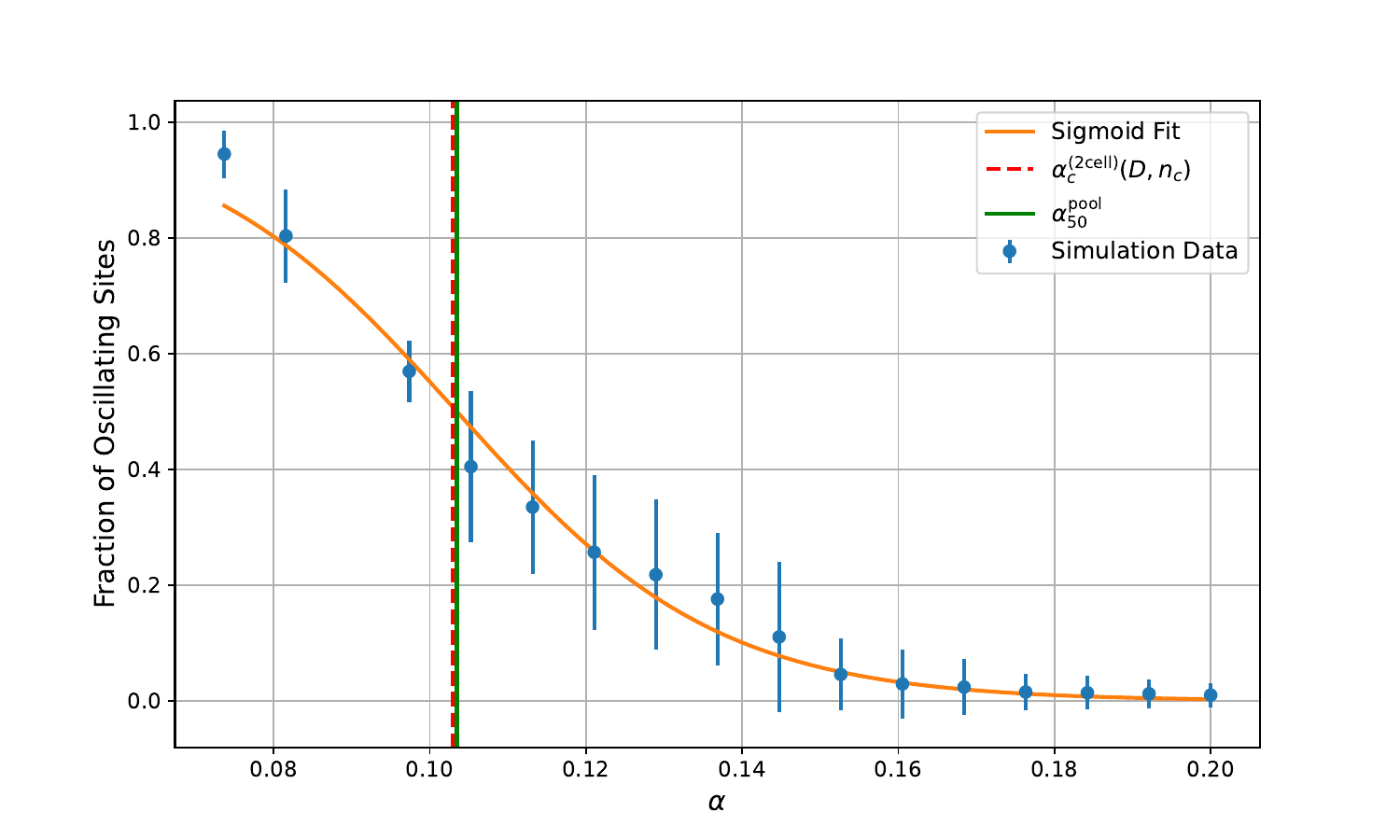}}
    \caption{Ensemble average sigmoid function $f_{\mathrm{osc}}(\alpha)$ at fixed $(D,n_c)$. The blue points and relative error bars have been obtained from 10 independent realisations of the dynamics of the system in Eq. \ref{eq:sys} in absence of noise. The orange curve is the logistic function obtained by fitting the ensemble average sigmoid function with the two-parameter sigmoid $f(\alpha)=1/\!\bigl(1+\exp[s(\alpha-\alpha_{50})]\bigr)$. 
Vertical green continuous and red dashed lines indicate respectively (i) the inflection $\alpha_{50}^{\mathrm{pool}}$ obtained from the  fit and (ii) the two-cell mean-field Hopf threshold $\alpha_c^{(2\mathrm{cell})}(D,n_c)$ computed by locating the fixed point as the solution of $\max\Re\lambda(\alpha,D,n_c)=0$. Parameters are $D=1$, $n_c=0.7$. }
    \label{fig:foscvsalpha}
\end{figure}

\section{Early–warning signals: simulation and feature extraction}
\label{sec_app:EWS}

This section supplements the early--warning analysis in the main text (Section \ref{sec:ews}). We first describe how we generate multivariate time series under slowly varying control gain, then detail the rolling metrics (spectral and non--spectral), the preprocessing applied in each analysis window, and how we turn metric trajectories into a binary classifier with ROC/AUC.
\subsection{Adiabatic time--series generation}

We simulate our $L\times L$ lattice (with $L=64$) of FitzHugh--Nagumo--type units coupled diffusively (Eq. \ref{eq:sys}). The control gain $\alpha(t)$ is prescribed to decrease \emph{slowly} in time according to Eq. \ref{eq:slow_alpha} with the following fixed choices:

\begin{itemize}
  \item \textbf{Numerical parameters.} Time step $\,\Delta t =0.1\,$s; initial burn--in $\;T_{\mathrm{burn}}=200\,$s (i.e.\ $N_{\mathrm{burn}}=T_{\mathrm{burn}}/\Delta t=2000$ steps) with control and $\alpha$ fixed; then $\alpha$ starts decreasing.
  \item \textbf{Ramp rate.} Linear ramp speed $\;\dot\alpha=-r$ with $r=5\times10^{-5}\,$.
\end{itemize}
We form two ensembles that serve as positive/negative classes for ROC evaluation:
\begin{itemize}
  \item \textbf{Transition ensemble.} During burn--in, we hold $\alpha$ at $\alpha_{\mathrm{start}}=\alpha_c+0.2$ and then \emph{decrease} it linearly down to $\alpha_{\mathrm{stop}}=\alpha_c$ at the fixed rate above. The ramp span is $\Delta\alpha=0.20$, hence
  \[
    T_{\mathrm{ramp}}^{(\mathrm{trans})}=\frac{\Delta\alpha}{r}
    \;=\;\frac{0.20}{5\times10^{-5}} \;=\; 4000\ \text{s}
    \quad\Rightarrow\quad N_{\mathrm{ramp}}^{(\mathrm{trans})}=40000\,\text{steps}.
  \]
  Total length: $N_{\mathrm{tot}}^{(\mathrm{trans})}=N_{\mathrm{burn}}+N_{\mathrm{ramp}}^{(\mathrm{trans})}=2000+40000=42000$ steps ($T_{\mathrm{tot}}^{(\mathrm{trans})}=4200$\,s).
  \item \textbf{Null (adiabatic) ensemble.} During burn--in, we hold $\alpha$ at $\alpha_{\mathrm{start}}=\alpha_c+0.6$ and then \emph{decrease} it linearly down to $\alpha_{\mathrm{stop}}=\alpha_c+0.4$ at the fixed rate above. The ramp span is $\Delta\alpha=0.20$ and so the length of the time series is the same as the transition one.
\end{itemize}
In both ensembles, the piecewise schedule can be written as
\[
  \alpha_k \;=\;
  \begin{cases}
    \alpha_{\mathrm{start}}, & 0 \le k < N_{\mathrm{burn}},\\[3pt]
    \alpha_{\mathrm{start}} + \dfrac{k-N_{\mathrm{burn}}}{N_{\mathrm{ramp}}}\,\bigl(\alpha_{\mathrm{stop}}-\alpha_{\mathrm{start}}\bigr), & N_{\mathrm{burn}} \le k < N_{\mathrm{burn}}+N_{\mathrm{ramp}},
  \end{cases}
\]
with $(\alpha_{\mathrm{start}},\alpha_{\mathrm{stop}})$ chosen as above for each ensemble. These two ensembles constitute the positive (approaching transition) and negative (quiescent) classes for ROC evaluation. Unless otherwise noted, we use additive observation noise $\sigma_V=0.1$ on $V_e$. We perform $N=100$ independent simulations per class and per $n_c$ value for ROC estimation (the coupling is kept fixed $D=1$).

\subsection{Rolling windows and preprocessing}
We summarize each multivariate trace into rolling statistics over windows of length $N_{\mathrm{win}}$ with stride $\Delta$. Unless noted otherwise we use
\[
N_{\mathrm{win}}=4096, \quad \Delta=512.
\]
All metrics are computed separately on three cell subsets
\[
S\in\{\text{controlled},\ \text{uncontrolled},\ \text{all}\},
\]
Within each window, we apply some or all of the following steps:
\begin{enumerate}
  \item \textbf{DC removal:} subtract the within--window mean from each cell's segment.
  \item \textbf{Linear detrending:} remove a best--fit line.
  \item \textbf{Hann taper:} multiply by $w_t=\tfrac{1}{2}\!\left(1-\cos\!\tfrac{2\pi t}{N_{\mathrm{win}}-1}\right)$, $t=0,\dots,N_{\mathrm{win}}-1$.
  \item \textbf{Variance standardization (spectral only):} for spectral metrics, each tapered segment is rescaled to unit standard deviation, ensuring normalized spectral amplitudes across windows; for time--domain power, we \emph{do not} standardize (to preserve amplitude growth).
\end{enumerate}

\subsection{Spectral metrics}
For each cell $i$ in a given window, let $x_{i,t}$ denote the preprocessed (demeaned, detrended, tapered, unit--variance) segment and $\widehat{X}_i[k]$ its one--sided discrete Fourier transform (positive frequencies). We define the (one-sided) periodogram (power spectral density estimate, normalized by $N^2$) as
\[
P_i[f_k] \;=\; \frac{2}{N^2}\,\bigl|\widehat{X}_i[k]\bigr|^2,\qquad f_k=\frac{k}{N\Delta t},\; k=1,\dots,\lfloor N/2\rfloor,
\]
excluding the DC component ($k=0$). The peak-power metric is then
\[
p_i^{\max} \;=\; \max_{k=1,\dots,\lfloor N/2\rfloor} P_i[f_k].
\]
We then obtain the following statistics across a subset $S$
\[
\text{avgpeak}_S=\frac{1}{|S|}\sum_{i\in S} p_i^{\max},
\qquad
\text{maxpeak}_S=\max_{i\in S} p_i^{\max}.
\]
We track both $\text{avgpeak}_S$ and $\text{maxpeak}_S$ over time for $S=\text{controlled, uncontrolled, all}$ (six spectral curves in total). These simple metrics are inspired by the work of \cite{bury2020detecting} where the authors apply spectral early warning signals (max peak of power spectrum) to one-dimensional systems.


\subsection{Time--domain power metrics}
Let $x_{i,t}$ denote the (demeaned, optionally detrended) segment in the window and $w_t$ the Hann taper. For each cell, we compute the mean square normalized by the taper energy of the window
\[
E_w=\frac{1}{N_{\mathrm{win}}}\sum_{t=0}^{N_{\mathrm{win}}-1} w_t^2,
\qquad
P_i=\frac{1}{E_w}\,\frac{1}{N_{\mathrm{win}}}\sum_{t=0}^{N_{\mathrm{win}}-1}\bigl(w_t\,x_{i,t}\bigr)^2.
\]
\emph{Crucially, no variance standardization is applied here}, so that changes in oscillation amplitude translate directly into rising power. The statistics per subset $S$ that we consider are
\[
\text{avgpower}_S=\frac{1}{|S|}\sum_{i\in S} P_i,
\qquad
\text{maxpower}_S=\max_{i\in S} P_i,
\]
again yielding six non--spectral curves (avg/max $\times$ three subsets).

\subsection{Turning metric curves into a classifier (ROC/AUC)}
For any rolling metric $M_S(t)$ (e.g.\ $\text{avgpower}_{\text{uncontrolled}}(t)$ or $\text{maxpeak}_{\text{all}}(t)$), we quantify whether it trends upwards by computing Kendall's rank correlation $\tau_K$ between $\{M_S(t)\}$ and time $t$ over the most recent $W$ windows:
\[
\tau_K \;=\; \frac{\#\text{concordant}-\#\text{discordant}}{\binom{W}{2}}\ \in[-1,1].
\]
We treat $\tau_K$ as a scalar score: thresholding $\tau_K$ yields a binary “approaching transition” decision. Comparing the score distributions from the transition and null ensembles produces ROC curves and associated AUC.

\subsection{Alternative metrics to the main text}
\begin{figure}[htbp]
  \centering

  \subfloat[\label{fig:maxauc}]{%
    \includegraphics[width=0.48\textwidth]{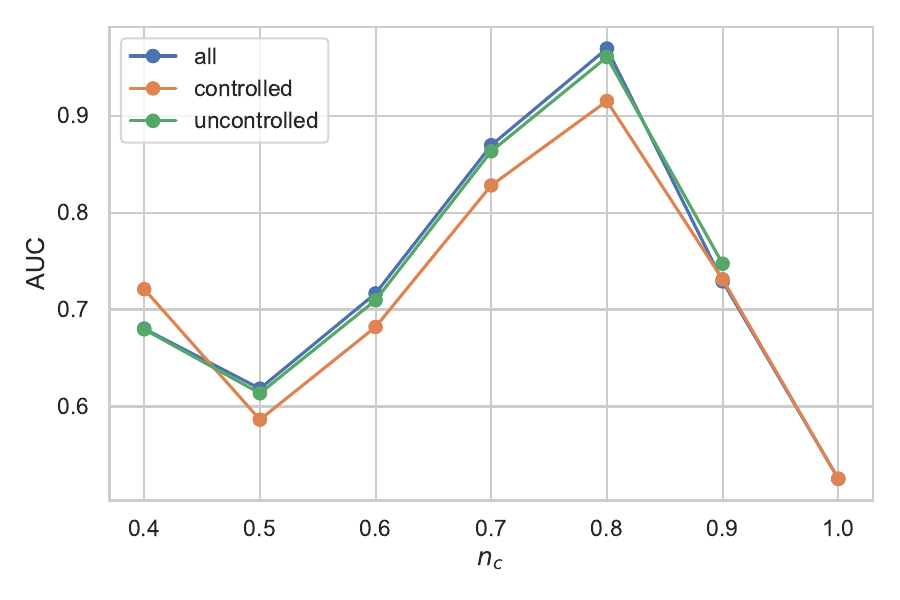}}
  \hfill
  \subfloat[\label{fig:maxfpr-at-tpr}]{%
    \includegraphics[width=0.48\textwidth]{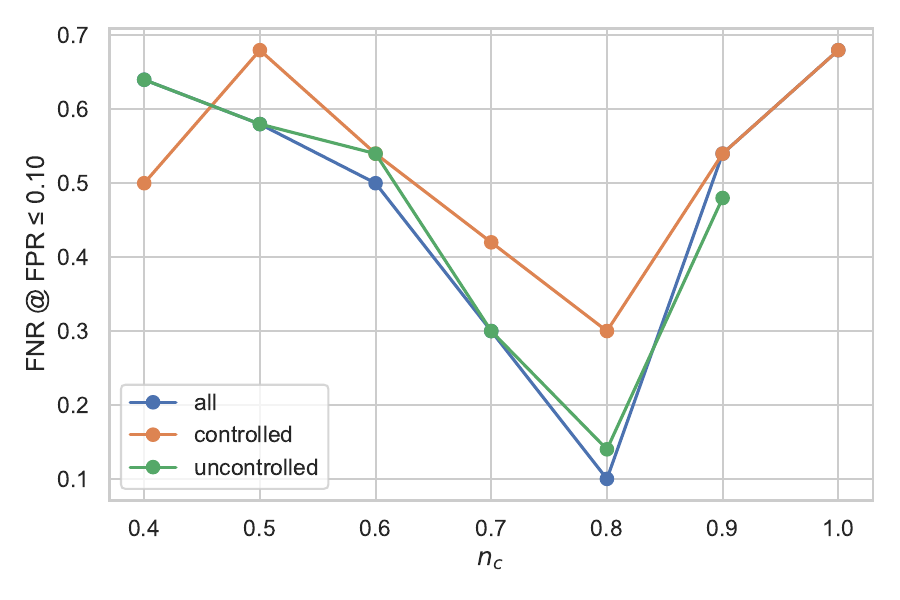}}

  \vspace{1.0em} 

  \subfloat[\label{figmax:tau_box_cont}]{%
    \includegraphics[width=0.48\textwidth]{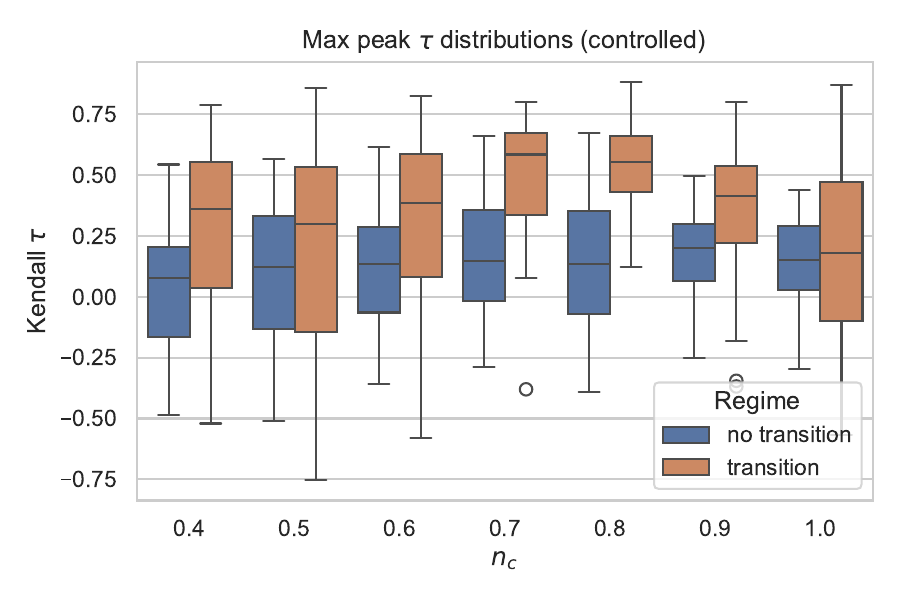}}
  \hfill
  \subfloat[\label{figmax:tau_box_uncont}]{%
    \includegraphics[width=0.48\textwidth]{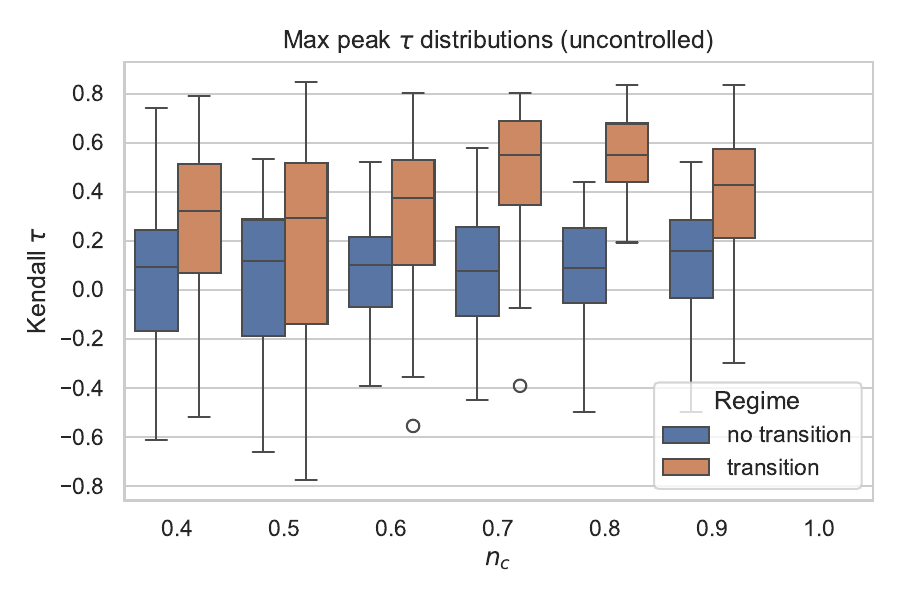}}

  \caption{Performance of the sentinel module treated as a binary classifier when using $\text{maxpeak}_S$ as metric.  
(a) Area under the ROC curve (AUC) as a function of $n_c$ for $D=1$.  
Each curve corresponds to a different subset of sites observed by the module: controlled cells, uncontrolled cells, or all cells. The AUC summarizes the trade-off between true positive rate and false positive rate when the decision threshold on $\tau_K$ is varied.  
(b) False negative rate (FNR) when the false positive rate (FPR) is constrained to be below 10\%.  
Here, $\mathrm{FNR} = \frac{\# \text{ imminent parturitions not detected}}{\# \text{ imminent parturitions}}$ and  
$\mathrm{FPR} = \frac{\# \text{ quiescent cases incorrectly flagged}}{\# \text{ quiescent cases}}$.  
The notation “FNR @ FPR” means that FNR is measured under the constraint that FPR $< 0.1$.  
(c–d) Box plots of the \emph{Kendall rank correlation coefficient} $\tau_K$ between $\text{maxpeak}_S$ and time $t$, computed over the most recent $W$ points of each time series.  
Panel (c) shows the box plots of $\tau_K$ when $\mathcal{S}$ is the set of controlled cells; panel (d) shows box plots of $\tau_K$ when $\mathcal{S}$ is the set of uncontrolled cells.  
In each panel, box plots are shown separately for the “approaching parturition” (positive) and “quiescent” (negative) classes, illustrating the degree of overlap between the two classes in $\tau_K$ space and thus how easily they can be separated by a threshold. Compare with Fig. \ref{fig:avgpower-eval}.
}
  \label{fig:maxpeak-eval}
\end{figure}

The main text focuses on $\text{avgpower}_S$ (see Fig. \ref{fig:avgpower-eval}) because it is robust and easy to interpret. The other mentioned metrics show qualitatively the same behaviour: detection performance improves with $n_c$, and performance is higher when the analysis is restricted to uncontrolled cells. An exception is the \emph{peak-power} metric $\mathrm{maxpeak}_S$ (the maximum of the periodogram across nonzero frequency bins, averaged over the observed sites), whose classifier performance exhibits a non-monotonic dependence on $n_c$ with a clear interior maximum near $n_c\approx 0.8$ (Fig.~\ref{fig:maxpeak-eval}).
We conjecture that this occurs because two performance drivers counterbalance each other: (i) higher $n_c$ enhances performance by amplifying the response to variations in $\alpha$, while (ii) uncontrolled cells contribute additional information. The max operator therefore performs best when only a few uncontrolled cells remain, provided they are sufficiently informative.

\end{document}